\documentclass[11pt]{articleKI}

\title{\bf Numerical behavior of the Keplerian Integral methods for
  initial orbit determination}

\author[1]{\'Oscar Rodr\'iguez\,\orcidlink{0000-0002-4545-5135}}
\author[1]{Giovanni F. Gronchi\,\orcidlink{0000-0003-1294-0633}}
\author[1]{Giulio Ba\`u\,\orcidlink{0000-0002-9857-0866}}
\author[2]{Robert Jedicke\,\orcidlink{0000-0001-7830-028X}}
\affil[1]{Dipartimento di Matematica, Universit\`a di Pisa, Italy}
\affil[2]{Institute for Astronomy, University of Hawai`i, USA}
\begin{document}
\maketitle

\begin{abstract}
We investigate the behaviour of two recent methods for the computation
of preliminary orbits.  These methods are based on the conservation
laws of Kepler's problem, and enable the linkage of very short arcs of
optical observations even when they are separated in time by a few
years.  Our analysis is performed using both synthetic and real data
of 822 main belt asteroids.  The differences between computed and true
orbital elements have been analysed for the true linkages, as well as
the occurrence of alternative solutions.  Some metrics have been
introduced to quantify the results, with the aim of discarding as many
of the false linkages as possible and keeping the vast majority of
true ones.  These numerical experiments provide thresholds for the
metrics which take advantage of the knowledge of the \emph{ground
  truth}: the values of these thresholds can be used in normal
operation mode, when we do not know the correct values of the orbital
elements and whether the linkages are true or false.
\end{abstract}



\section{Introduction}

Modern asteroid surveys, like Pan-STARRS \citep{Denneau2013-MOPS} and
Catalina \citep{Catalina}, collect a large number of optical
observations that are grouped in {\em very short arcs} (VSA), also
called {\em tracklets}. The information contained in a VSA is usually
not sufficient to compute a reliable least squares orbit with
classical methods \citep{laplace,lagrange,gauss1809}.  In this case,
we can try to compute an orbit by combining the data from two or more
VSAs under the assumption that they belong to the same celestial body.
This operation is called {\em linkage}, see \citet[Chap. 7]{mg10}.

Among the existing linkage algorithms, here we consider the so-called
Keplerian integrals (KI) methods.  These methods provide a polynomial
equation for the computation of a preliminary orbit from the
conservation laws of Kepler's dynamics, i.e. angular momentum, the
Laplace-Lenz vector and energy.  The conservation of angular momentum
and energy were already used for this problem in \cite{TaffHall1977,
  Taff1984}, however the authors did not fully exploit the algebraic
character of the resulting equations.  More recently, in a series of
papers \cite{gdm10,gfd11,gbm15} derived from the Keplerian
conservation laws polynomial equations of degree 48, 20 and 9,
respectively, for the linkage of two VSAs.  In \cite{gbm17} the
authors showed that the polynomial of degree 9 introduced in
\cite{gbm15} is optimal in some sense, and derived an equation of
degree 8 for the linkage of three VSAs.

In this work we analyse the numerical behaviour of two KI methods,
here denoted by \texttt{link2} and \texttt{link3}.  The first method,
introduced in \cite{gbm15}, uses two VSAs and a suitable combination
of all the conserved quantities.  The second, presented in
\cite{gbm17}, employs three VSAs and the conservation of the angular
momentum only.  These algorithms allow for a fast computation of
orbits, but their sensitivity to astrometric errors has not yet been
systematically analysed.  Understanding the performance of these
methods is crucial when working with large databases of VSAs such as
the isolated tracklet file
(ITF)\footnote{\url{http://www.minorplanetcenter.net/iau/ITF/itf.txt.gz}}
which is available from the Minor Planet Center (MPC).  The ITF is an
ever changing list of unlinked tracklets, mainly provided by
Pan-STARRS1 and Catalina, with more than 4 and 2 million observations,
respectively.  However, there are also observations from many other
telescopes, thus the quality of the data in the ITF is heterogeneous.
For this reason, we study the sensitivity of \texttt{link2} and
\texttt{link3} to different levels of astrometric errors.  During the
last years some effort has been made to compute orbits with the ITF
data \citep{Sansaturio2012,HelioLinC2018, WerykITF2020} and the size
of this file has been considerably reduced.  \cite{HelioLinC2018} used
a tracklet clustering technique to define an algorithm with complexity
$\mathcal{O}(N\mathrm{log}N)$, where $N$ is the total number of
tracklets.  \cite{Sansaturio2012} employed an identification algorithm
of attribution type that took into account the higher apparent rates
of motion of NEAs.

From the results of our study, we think that the KI methods can be
efficient tools for initial orbit determination, complementary to the
existing ones. In fact, as opposite to the other algorithms, they are
able to link tracklets that are separated in time even by a few years.

The structure of this paper is the following. In Section
\ref{sec:KImethods} we present the \texttt{link2} and \texttt{link3}
methods and the indicators that we use to analyse the quality of the
solutions.  In Section \ref{sec:datasets} we explain the generation of
the test data sets. Section \ref{sec:synthetic} is devoted to the
analysis of the numerical behaviour of \texttt{link2} and
\texttt{link3} when applied to synthetic data.  In Section
\ref{sec:realdata} we apply the two KI methods to real data and we
compare their performance to that obtained in
Section~\ref{sec:synthetic}.  Finally, in Section \ref{sec:ITF} we
estimate the efficiency of \texttt{link2} in producing correct
linkages under simplistic assumptions: in particular, we estimate the
number of sets of 4 tracklets (correctly or incorrectly associated)
that need to be examined as a function of the probability of
identifying correct associations.

\section{The KI methods}
\label{sec:KImethods}

Let us consider a set of $m\geq2$ optical observations of a celestial
body $\{(\alpha_i,\delta_i)\,|\, i =1,...,m\}$, where $\alpha_i,
\delta_i$ are right ascension and declination at epochs $t_i$,
$i=1,...,m$, that are close in time to each other.  From these data we
can compute the \emph{attributable} vector
\[
\bm{\mathcal{A}} = (\alpha,\delta,\dot{\alpha},\dot{\delta}),
\]
representing the angular position and angular rate of the body at the
mean time $\bar{t} = \frac{1}{m}\sum_{i=1}^m t_i$. We denote by
$\bm{r}$ and $\bm{q}$ the heliocentric positions of the asteroid and
the observer at time $\bar{t}$.  In this way, the topocentric position
of the asteroid is given by $\bm{\rho} = \bm{r}-\bm{q}$.  Setting
$\rho = |\bm{\rho}|$, we can write $\bm{\rho} = \rho {\bf e}^\rho$,
where
\[
{\bf e}^\rho = (\cos\alpha\cos\delta,\sin\alpha\cos\delta,\sin\delta),
\]
is a known vector, the \emph{line of sight}.

The Keplerian integrals (KI) methods exploit the conservation laws of
Kepler's dynamics for the linkage.  The conserved quantities are
\begin{equation}
  \bm{c} = \bm{r} \times \dot{\bm{r}},\qquad
  \mathcal{E} = \frac{1}{2}|\dot{\bm{r}}|^2 -\frac{\mu}{|\bm{r}|},\qquad 
  {\bf L} = \frac{1}{\mu}\dot{\bm{r}}\times\bm{c} - \frac{\bm{r}}{|\bm{r}|},
  \label{eq:kepint}
\end{equation}
which correspond to the angular momentum, energy, and Laplace-Lenz
vector of the orbit, respectively.

Given an attributable $\bm{\mathcal{A}}$ at the epoch $\bar{t}$, the
quantities \eqref{eq:kepint} can be written as algebraic functions of
the unknowns $\rho, \dot{\rho}$, with coefficients depending on the
attributable $\bm{\mathcal{A}}$ and on the heliocentric position
$\bm{q}$ and velocity $\dot{\bm{q}}$ of the observer.

\subsection{The \texttt{link2} algorithm}

We briefly recall the algorithm introduced in \cite{gbm15} for the
linkage of two VSAs. Given two attributables of the same object
$\bm{\mathcal{A}}_1, \bm{\mathcal{A}}_2$ at the epochs $\bar{t}_1,
\bar{t}_2$, we consider the system
\begin{equation}
  \bm{c}_1 =\bm{c}_2, \qquad [\mu({\bf L}_1-{\bf L}_2) -
  (\mathcal{E}_1\bm{r}_1-\mathcal{E}_2\bm{r}_2)]\times(\bm{r}_1-\bm{r}_2) = \bm{0},
  \label{eq:KIsys}
\end{equation}
where the subscripts refer to the two epochs and we assume 2-body
Keplerian motion between the two times.  System \eqref{eq:KIsys} is
polynomial, is composed of 6 equations in 4 unknowns $(\rho_1, \rho_2,
\dot{\rho}_1, \dot{\rho}_2)$, and is therefore over-determined.
Nevertheless, we can show that this system is consistent, i.e. it
always has solutions, at least in the complex field, even when the two
tracklets belong to different objects.

By elimination of variables, system \eqref{eq:KIsys} leads to a
polynomial equation of degree 9 in one of the two unknown topocentric
ranges ($\rho_1$ or $\rho_2$).

From the positive roots of this polynomial we obtain quadruples of
solutions of \eqref{eq:KIsys}. These correspond to pairs of Keplerian
orbits at epochs $\tilde{t}_1, \tilde{t}_2$, obtained by applying
aberration correction to $\bar{t}_1$, $\bar{t}_2$.

\subsubsection{The \texorpdfstring{$\chi_4$}{chi4} norm}
\label{sec:chi4}

The preliminary orbits computed with the \texttt{link2} algorithm have
an associated covariance matrix. Consider a preliminary orbit at
epoch $\tilde{t}_1$. This orbit and its covariance are propagated to
the epoch $\bar{t}_2$ of the second attributable $\bm{\mathcal{A}}_2$,
with covariance matrix $\Gamma_2$. Then, we can compute a propagated
attributable $\bm{\mathcal{A}}_p$, with marginal covariance matrix
$\Gamma_{p}$, see \citep[Chap.7]{mg10}, \citep{ident4}.  We quantify
the difference between the two attributables $\bm{\mathcal{A}}_p$ and
$\bm{\mathcal{A}}_2$ by the $\chi_4$ norm,
\begin{equation}
  \chi_4= \sqrt{(\bm{\mathcal{A}}_p - \bm{\mathcal{A}}_2)\cdot
  \left [C_2-C_2\,\Gamma_0\,C_2\right]\,(\bm{\mathcal{A}}_p - \bm{\mathcal{A}}_2)^t},
  \label{eq:chi4}
\end{equation}
where
\[
C_2=\Gamma_2^{-1},\qquad \Gamma_0 = C_0^{-1},
\]
with
\[
C_0 = C_2 + C_p, \qquad C_p = \Gamma_p^{-1}.
\]
We will show that the $\chi_4$ norm can be used to select good
solutions from \texttt{link2} using a threshold for the acceptable
solutions.

\subsection{The \texttt{link3} algorithm}

Given three attibutables $\bm{\mathcal{A}}_1$, $\bm{\mathcal{A}}_2$,
$\bm{\mathcal{A}}_3$ of an asteroid, one or more preliminary orbits
can be computed by imposing the conservation of the angular momentum
only:
\begin{equation}
  \bm{c}_1 = \bm{c}_2,
  \qquad
  \bm{c}_2 = \bm{c}_3.
  \label{eq:angmom}
\end{equation}
System \eqref{eq:angmom} gives us 6 scalar equations in the 6 unknowns
$\rho_1$, $\dot{\rho}_1$, $\rho_2$, $\dot{\rho}_2$, $\rho_3$,
$\dot{\rho}_3$. Let us define
\[
\bm{D} = \bm{q} \times {\bf e}^\rho.
\]
If relation
\[
\bm{D}_1\times\bm{D}_2\cdot\bm{D}_3 \neq 0,
\label{D1D2D3}
\]
holds, then equations \eqref{eq:angmom} are equivalent to
\[\label{eq:angmomLK3}
\begin{aligned}
  0 &= (\bm{c}_i-\bm{c}_j)\cdot\bm{D}_i\times\bm{D}_j,\\
  0 &= (\bm{c}_i-\bm{c}_j)\cdot\bm{D}_i\times(\bm{D}_i\times\bm{D}_j), 
\end{aligned}
\]
with $(i,j) = (1,2),(2,3),(3,1)$. By elimination of variables, as
explained in \cite{gbm17}, we can write a polynomial equation of
degree 8 in the $\rho_2$ unknown only, and then reconstruct the values
of the other variables. Note that we can always discard one of the
solutions, because it corresponds to a \emph{straight line} solution,
with zero angular momentum.

\subsubsection{The \emph{star} norm}
\label{sec:starmetric}

The \texttt{link3} method only imposes the equality of the angular
momentum $\bm{c}$ at the three different times.  Note that, given a
Keplerian orbit defined by the orbital elements
$(a,e,i,\Omega,\omega,\ell)$, we can express its angular momentum as
\[
\bm{c} = \sqrt{\mu a(1-e^2)}\left(\sin\Omega\sin i,\, -\cos\Omega\sin i,\, \cos i\right),
\]
where $\mu$ is the gravitational parameter. Therefore, while
$\Omega$, $i$, and $c^2 = \mu a(1-e^2)$ take the same values at the
three epochs, the orbital elements $\omega$, $\ell$, $a$, and $e$
might be different. In light of this consideration, we introduce a
norm that accounts for such differences.

Given a triplet of attributables $\bm{A} =
(\bm{\mathcal{A}}_1,\bm{\mathcal{A}}_2,\bm{\mathcal{A}}_3)$ with
covariance matrices $\Gamma_{\bm{\mathcal{A}}_j}$ we consider the
difference
\[
\begin{aligned}
  \bm{\Delta}_{jk} = \Big(& a_j - a_k,\,
  \big(\omega_j - \omega_k +  \pi\big) (\mathrm{mod}\, 2\pi) - \pi,\, \\
  & \left(\ell_j - \left[\ell_k + n(a_j)(\tilde{t}_j-\tilde{t}_k)\right] +
  \pi\right)(\mathrm{mod}\, 2\pi) - \pi\Big),
\end{aligned}
\]
where $n(a)=\sqrt{\mu/a^3}$ is the mean motion and $j,k \in\{1,2,3\}$
refer to the quantities at the three times. We introduce the vector
$\bm{\Delta} = (\bm{\Delta}_{12},\bm{\Delta}_{32})$ and the matrix
\[
\Gamma_{\bm{A}} = 
\begin{pmatrix}
  \Gamma_{\bm{\mathcal{A}}_1} & 0 & 0 \\
  0 & \Gamma_{\bm{\mathcal{A}}_2} & 0 \\
  0 & 0 &\Gamma_{\bm{\mathcal{A}}_3}  \\
\end{pmatrix}.
\]
Setting
\[
\Gamma_{\bm{\Delta}} = \frac{\partial \bm{\Delta}}{\partial \bm{A}} \Gamma_{\bm{A}}
\left[\frac{\partial \bm{\Delta}}{\partial \bm{A}}\right]^t,\qquad C_{\bm{\Delta}} =
\Gamma_{\bm{\Delta}}^{-1},
\]
we define the $\star$ norm as 
\begin{equation}
  \bm{\Delta}_\star = \sqrt{\bm{\Delta}C_{\bm{\Delta}}\bm{\Delta}^t}.
\end{equation}
More details about the computation of this norm can be found in
\cite{gbm17}.

\subsection{The \emph{rms} norm}
\label{sec:rms}

Another metric for the quality of preliminary orbits is:
\begin{equation} \label{eq:rms}
  rms = \sqrt{\frac{1}{n}\sum_{i=1}^n \Big[ 
  \Delta_{\alpha_i}^2\cos^2\delta_i + \Delta_{\delta_i}^2\Big]},
\end{equation}
where $\Delta_{\alpha_i} = \alpha_i-\alpha(\bar{t}_i)$,
$\Delta_{\delta_i} = \delta_i-\delta(\bar{t}_i)$ are the residuals of
the observations, with $\alpha(\bar{t}_i), \delta(\bar{t}_i)$ coming
from a 2-body propagation of the preliminary orbit.

\subsection{Least squares norm}
\label{sec:LS-rms}

After computing a preliminary orbit with \texttt{link2} or
\texttt{link3}, a differential corrections scheme \cite[Chap. 5]{mg10}
can be used to determine a least squares orbit.\footnote{We used the
  \texttt{fdiff\_cor} routine of the OrbFit package for the
  differential corrections (\url{http://adams.dm.unipi.it/orbfit/)}.}
If this scheme is successful, we can define the following metric:
\begin{equation} \label{eq:LSrms}
  R_{LS} = 
  \sqrt{\frac{1}{n}\sum_{i=1}^n \Big[w_i\Big( 
  \Delta_{\alpha_i}^2\cos^2\delta_i + \Delta_{\delta_i}^2\Big)\Big]},
\end{equation}
with $\alpha(\bar{t}_i), \delta(\bar{t}_i)$ coming from the full
$n$-body propagation of the least squares orbit. Here we allow
different weights $w_i$ for the observations.

Differential corrections are computationally more expensive than the
algorithms for the determination of a preliminary orbit, and the
iterative process must begin with a `good' orbit close to reality.  If
the preliminary solution computed with \texttt{link2} or
\texttt{link3} is not good enough then differential corrections will
not converge or will converge to a wrong solution.

\section{The data sets}
\label{sec:datasets}

\begin{figure}[!ht]
  \centering
  \includegraphics[trim={14mm 0 14mm 0},clip,width=0.66\textwidth]{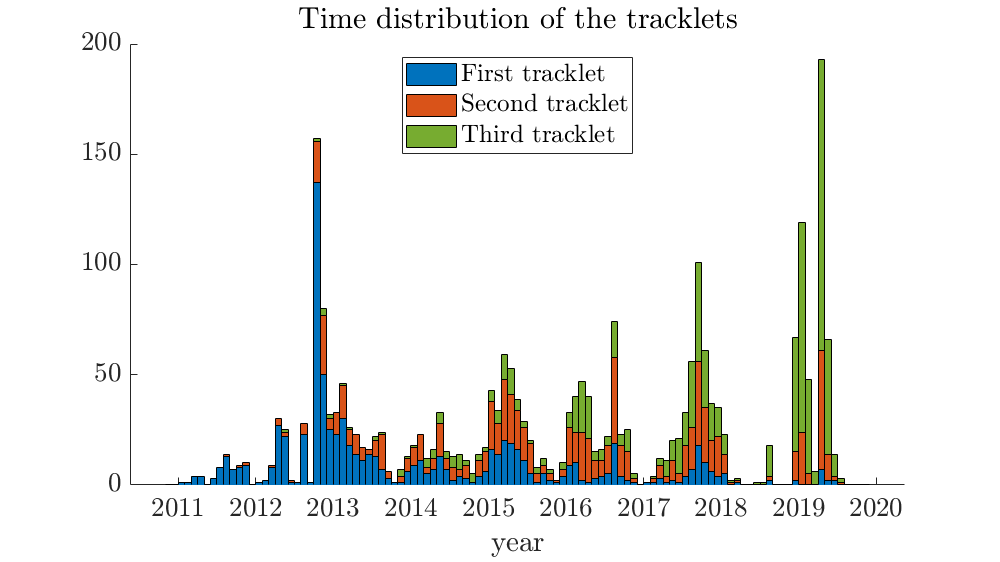}
  \caption{Time distribution of the tracklets in our \emph{matched
      real} data sample.}
  \label{fig:trackletTimeDist}
\end{figure}

We generated our test data sets from real data submitted by
Pan-STARRS1 \citep{Denneau2013-MOPS} to the MPC over the 8.5 year
period from 2011-01-30 through 2019-07-28 inclusive (see Figure
\ref{fig:trackletTimeDist}).  This technique ensures that we are using
consistent data from a single survey with a real observation cadence
within a night and across many years. We restricted the test data to
main belt objects because they dominate the statistics of any asteroid
survey and extracted triplets of random tracklets corresponding to the
same object for 822 asteroids.

\begin{figure}[!ht]
  \centering
  \includegraphics[trim={14mm 0 14mm 0},clip,width=0.66\textwidth]{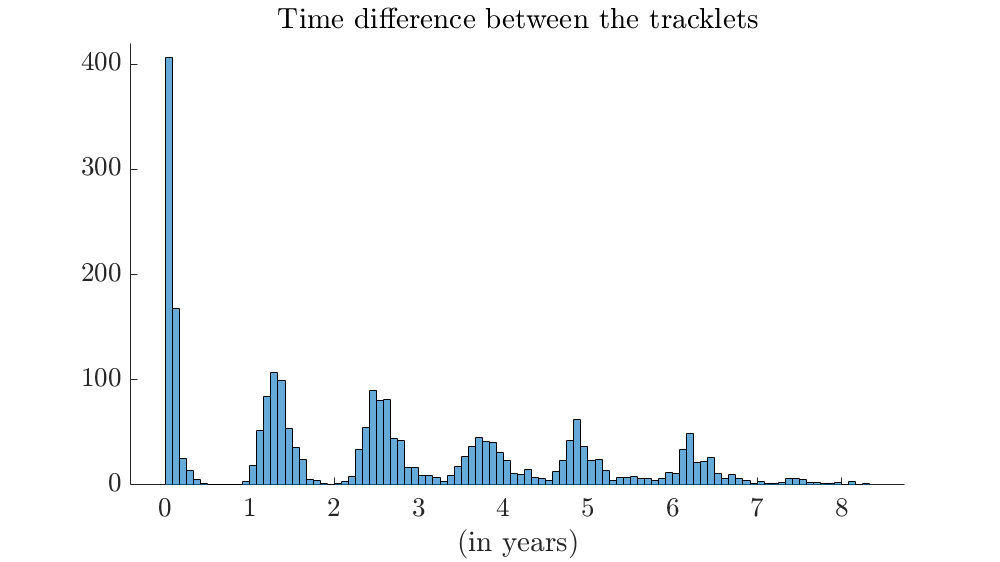}
  \caption{Time difference between pairs of tracklets for the same
    object.}
  \label{fig:trackletTimeDifference}
\end{figure}

We also required that each of the tracklets must have $\ge3$
detections. These observations are our \emph{matched real} data which
include all the vagaries of an actual operational survey including the
requirements that the observations be acquired at night, when the
telescope is operational, when the sky is clear, scheduling issues,
etc. The average time between any pair of tracklets corresponding to
the same object is about 2.6 years. We selected this data set in order
to test the capability of the algorithms to link tracklets with
different time separation (see Figure
\ref{fig:trackletTimeDifference}).

We then generated synthetic data sets for the same objects using the
nominal orbit for each object as reported by the MPC and the actual
times of observations from the \emph{matched real} Pan-STARRS1 data.
We did so using a simple 2-body calculation and a full $n$-body
integration with all the planets and major asteroids as implemented in
OpenOrb \citep{Granvik2009-OpenOrb}.  In both the 2-body and $n$-body
cases we generated synthetic data sets in which a random error was
introduced into the astrometric observations by generating a random
offset from the calculated positions according to a 2d Gaussian with
standard deviations ($\sigma$) of 0.1$\arcsec$, 0.2$\arcsec$,
0.5$\arcsec$, and 1.0$\arcsec$.

The synthetic data set with no introduced error allows us to determine
our algorithm's performance on perfect data.  The other data sets
allow us to characterise how the algorithms's performance degrades
with increasing astrometric error typical of other and historical
asteroid surveys with our eventual goal of applying \texttt{link2} and
\texttt{link3} to the MPC's ITF.

Note that the data sets used for \texttt{link2} are obtained from
those of \texttt{link3} by randomly selecting two tracklets per object
from the three available (i.e. 1644 tracklets for each data set), and
the same choice is maintained throughout all the numerical tests.

\section{Testing the KI methods with synthetic data}
\label{sec:synthetic}

In this section we characterise the performance of \texttt{link2} and
\texttt{link3} in terms of their sensitivity to astrometric error.

We first consider the case of \emph{true} linkages, i.e. linkages of
tracklets belonging to the same object.  In Section ~\ref{sec:TLinks}
we shall investigate how many \emph{false} linkages (i.e. linkages of
tracklets belonging to different objects) produce solutions.

\texttt{link3} is applied to the triplet of tracklets generated for
each object while \texttt{link2} is applied to two tracklets as
explained in Section~\ref{sec:datasets}. In this way, we determine the
percentage of cases for which we are able to find solutions with each
method, and investigate how this percentage evolves as we increase the
error in the simulated data. For the moment, we are not concerned with
the quality of the solutions.

\begin{figure}[ht!]
  \centering
  \includegraphics[width=0.66\textwidth]{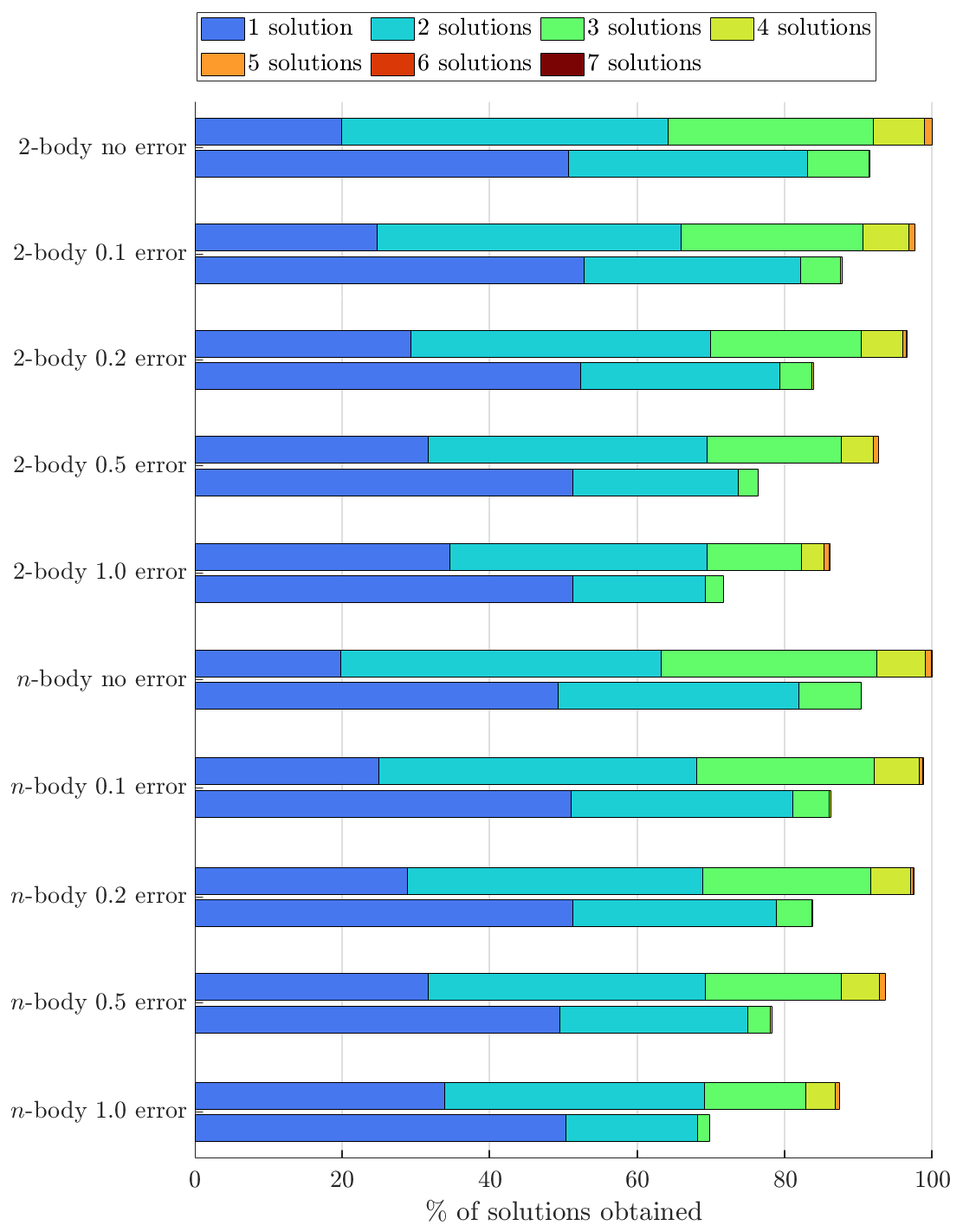}
  \caption{Percentage of linkages and multiplicity of the solutions
    obtained with \texttt{link2} (above) and \texttt{link3} (below)
    for each of the 10 synthetic data sets.}
  \label{fig:lk2vslk3solutions}
\end{figure}

The percentage of linkages found, out of all the possible true ones,
is high with both methods (see Figure~\ref{fig:lk2vslk3solutions}).
\texttt{link2} recovers a higher percentage of linkages than
\texttt{link3}. In particular, it allows us to find more than 86\% of
the linkages when the astrometric error is large, and more than 95\%
when the error is small, reaching 100\% in the ideal case of
observations with no error with both dynamical models (2-body and
$n$-body). \texttt{link3} is affected by astrometric error in a more
severe way, it identifies only about 70\% of the linkages with a large
error in the observations, and does not achieve the same high
percentages of linkages found by \texttt{link2} even when the error is
small or null. Most of the linkages recovered with \texttt{link2}
provide more than one solution. This also occurs with \texttt{link3}
but the percentage is lower.

\begin{figure*}
  \centering
  \includegraphics[width=0.84\textwidth]{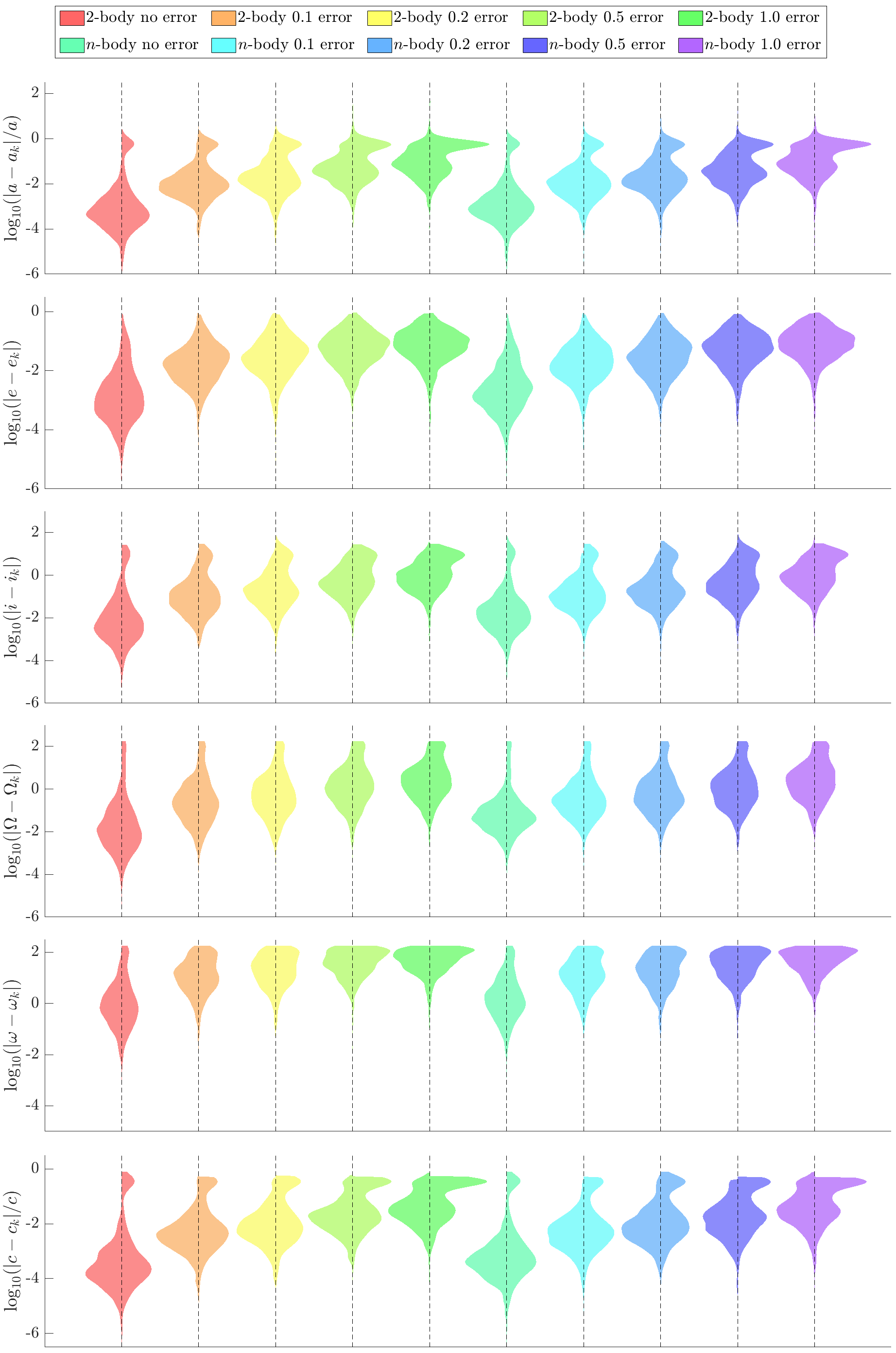}
  \caption{PDF of the logarithm of the error in $a$, $e$, $i$,
    $\Omega$, $\omega$, and $c$ (top to bottom) for each synthetic
    data set obtained from the solutions of \texttt{link2} (left of
    the dashed black line) and \texttt{link3} (right of the dashed
    black line).}
  \label{fig:lk2vslk3true}
\end{figure*}

A natural question that arises at this point is whether the quality of
the solutions obtained is good. To answer this question we
investigate the errors in the orbital elements $(a,e,i,\Omega,\omega)$
and in the angular momentum $c$ computed from the output of
\texttt{link2} and \texttt{link3}.  In Figure~\ref{fig:lk2vslk3true}
we plot the probability density function (PDF) of the logarithm of the
error for each component and for the different data sets.  This figure
was produced by choosing always the solution with the lowest value of
the orbit comparison criterion defined in
\cite{sh1963,Dcriterion2000}, hereafter simply $D$.  The quality of
the solutions is better using \texttt{link2} than \texttt{link3}.  The
PDF obtained with \texttt{link2} has a single local maximum and, as
expected, the performance deteriorates for larger astrometric errors.
The computed values of some elements are closer to the true ones than
others: the orbital plane (defined by $i$ and $\Omega$) is much better
determined than the perihelion position (given by $\omega$).
Furthermore, when the error is greater than 0.5$\arcsec$ the value of
$\omega$ is completely wrong.

In Figure~\ref{fig:lk2vslk3true} the PDF often presents two local
maxima with \texttt{link3}, and the second maximum increases with the
astrometric error. In this case we have a large number of false
solutions that need to be discarded \emph{a posteriori}. Like
\texttt{link2}, \texttt{link3} determines some orbital elements better
than others.  Considering also that \texttt{link3} recovers fewer
solutions than \texttt{link2} (see
Figure~\ref{fig:lk2vslk3solutions}), we conclude that it is preferable
to use the latter.

\begin{figure}[!ht]
  \centering
  \includegraphics[width=0.66\textwidth]{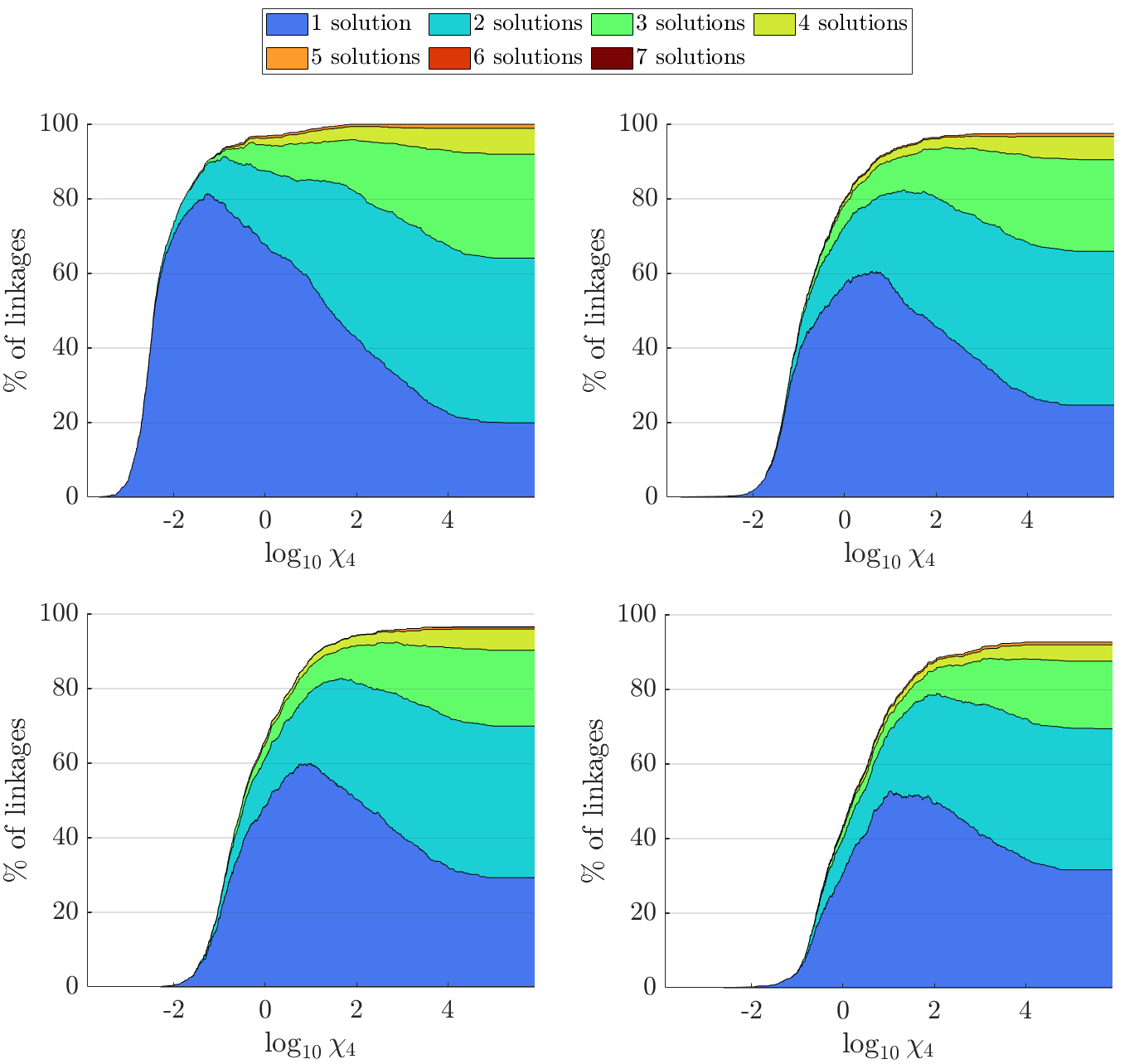}
  \caption{Evolution of the percentage of linkages recovered with a
    different number of solutions as a function of the $\chi_4$ norm.
    The synthetic data are generated by a 2-body propagation with no
    error, and 0.1$\arcsec$, 0.2$\arcsec$, 0.5$\arcsec$ error (from
    left to right and top to bottom).}
  \label{fig:lk2nomvsnumsol}
\end{figure}

As we have mentioned, both methods often give multiple solutions.
Some of them can be discarded using indicators such as the $\chi_4$
with \texttt{link2} or the $\Delta_\star$ with \texttt{link3} (see
Sections~\ref{sec:chi4}, \ref{sec:starmetric}). These norms will play
a crucial role to rule out false linkages. In
Figure~\ref{fig:lk2nomvsnumsol} we see how the percentage of linkages
with a different number of accepted solutions changes as the value of
the acceptance threshold of the $\chi_4$ increases. We note that by
choosing the value of this threshold corresponding to the dark blue
peak, we loose only $\approx 10\%$ of the possible linkages and most
of the accepted linkages give only one solution. This phenomenon is
shown by all data sets, being more evident with a small or zero error
(see for example the left plot of Figure~\ref{fig:lk2nomvsnumsol}).

\begin{figure}
  \centering
  \includegraphics[width=0.33\textwidth]{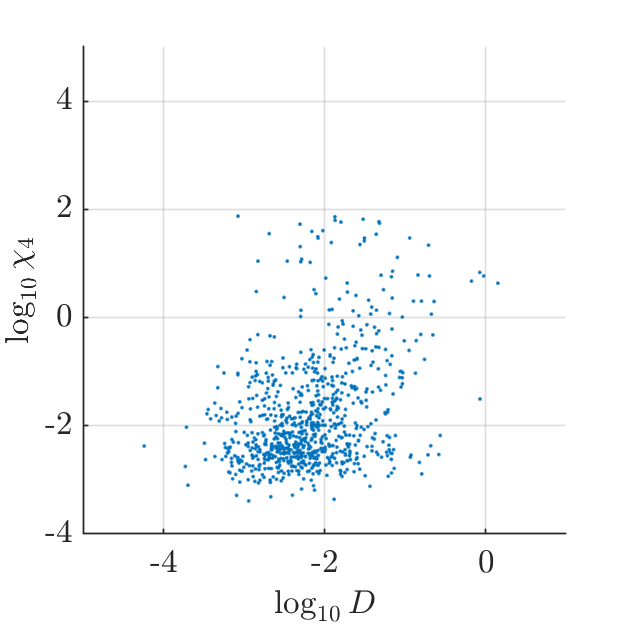}
  \includegraphics[width=0.33\textwidth]{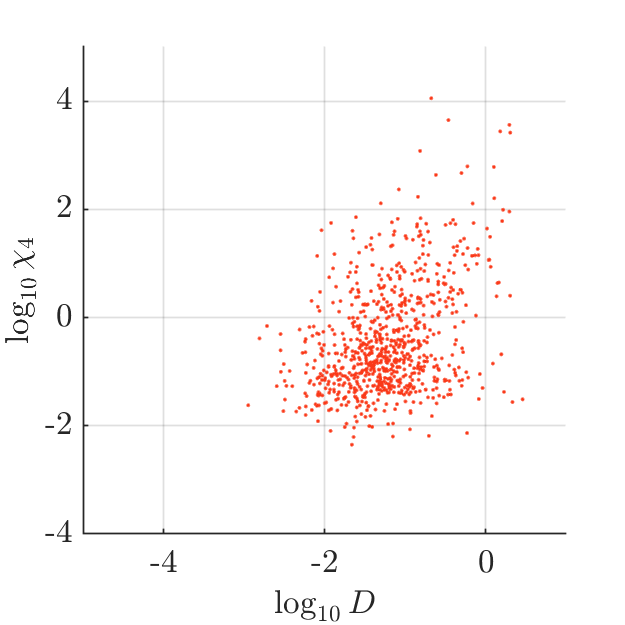}\\
  \includegraphics[width=0.33\textwidth]{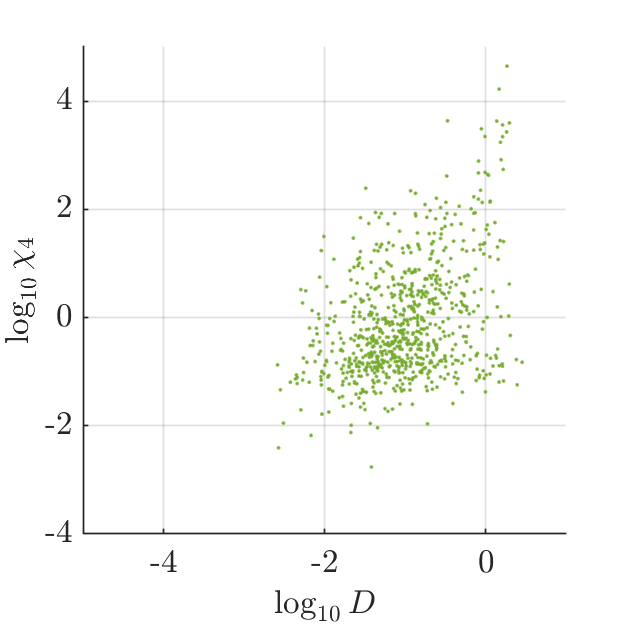}
  \includegraphics[width=0.33\textwidth]{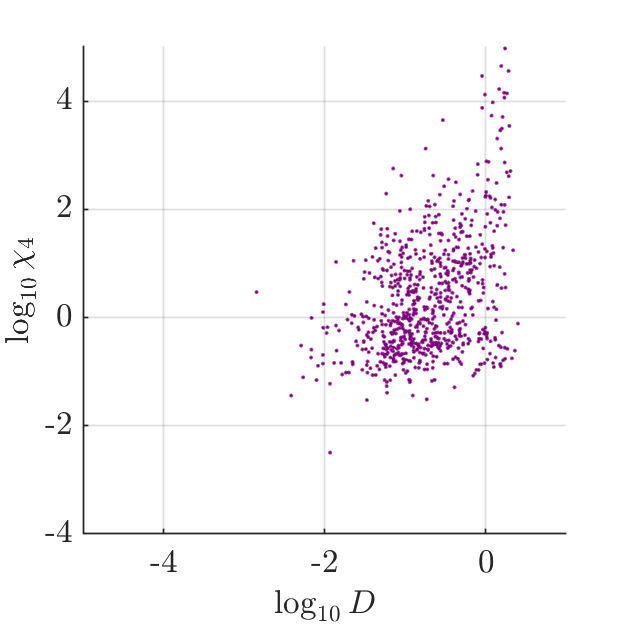}
  \caption{Values of $\chi_4$ vs $D$ in a log-log plot for
    \texttt{link2} using the synthetic data generated by an $n$-body
    propagation with no error, and 0.1$\arcsec$, 0.2$\arcsec$,
    0.5$\arcsec$ error (from left to right and top to bottom).}
  \label{fig:DcritVSNormLK2}
\end{figure}

\begin{figure}[ht!]
  \centering
  \includegraphics[width=0.33\textwidth]{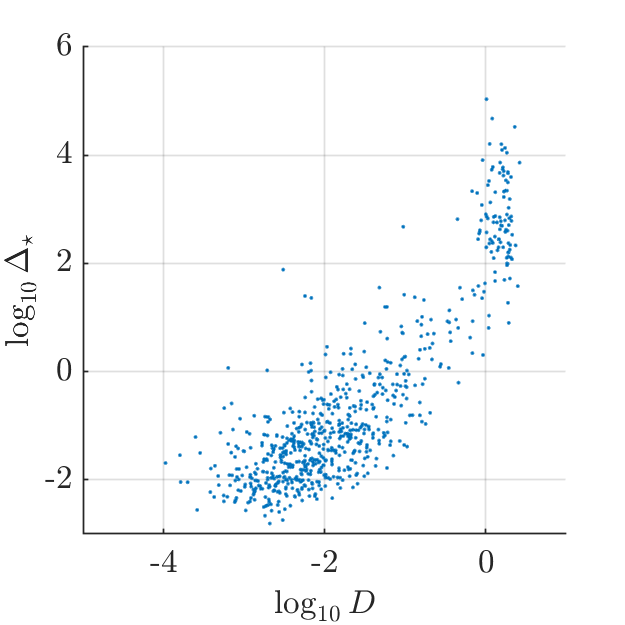}
  \includegraphics[width=0.33\textwidth]{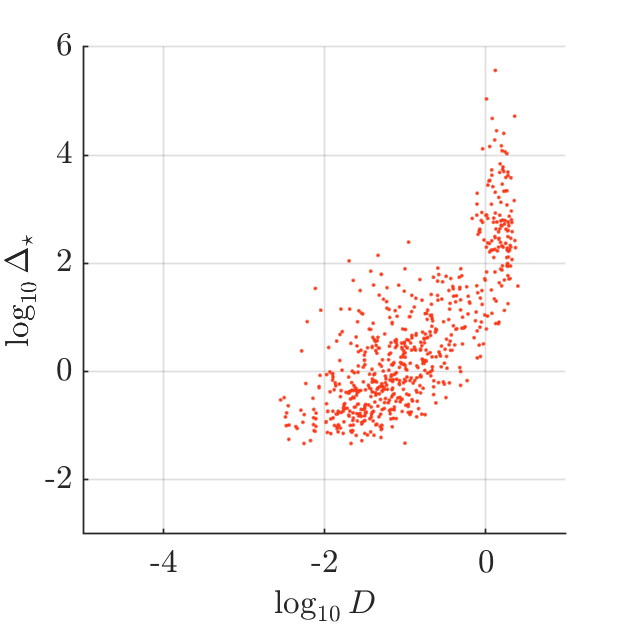}\\
  \includegraphics[width=0.33\textwidth]{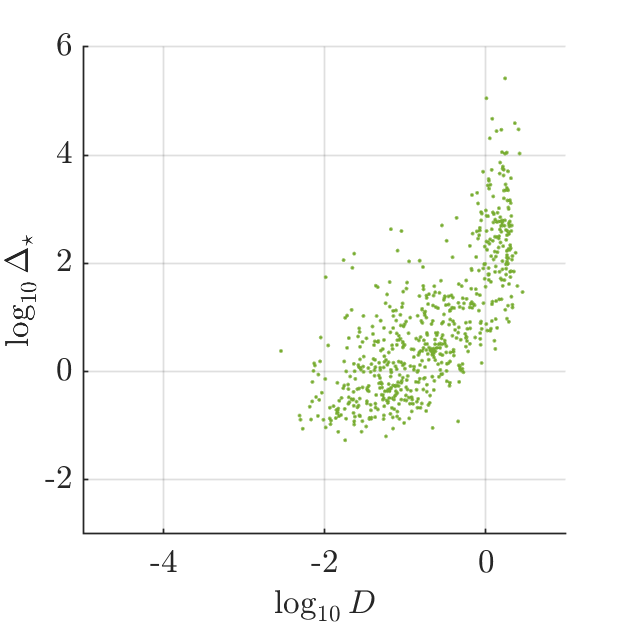}
  \includegraphics[width=0.33\textwidth]{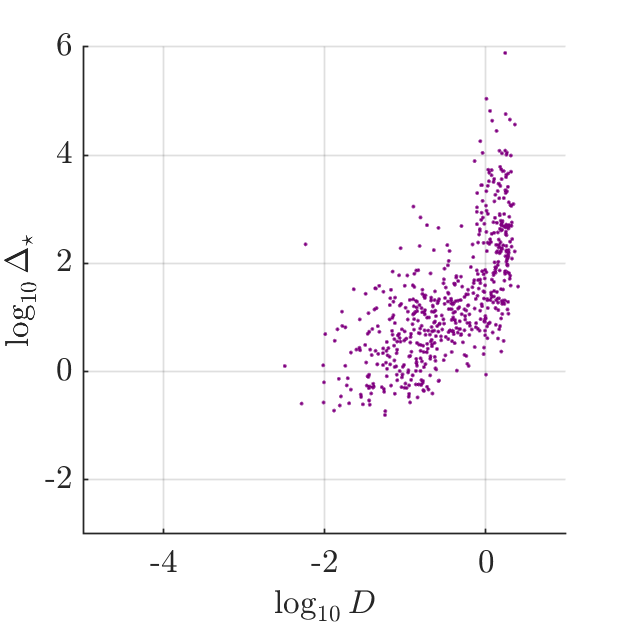}
  \caption{Values of $\Delta_\star$ vs $D$ in a log-log plot for
    \texttt{link3} using the synthetic data generated by an $n$-body
    propagation with no error, and 0.1$\arcsec$, 0.2$\arcsec$,
    0.5$\arcsec$ error (from left to right and top to bottom).}
  \label{fig:DcritVSNormLK3}
\end{figure}

The values of the $\chi_4$ and $\Delta_\star$ norms are important to
evaluate the quality of the solutions (see
Figures~\ref{fig:DcritVSNormLK2} and \ref{fig:DcritVSNormLK3}).  It is
common to identify two orbits when $D\leq 0.2$, see
\citep{sh1963,schunova2012}. In Figure~\ref{fig:DcritVSNormLK2} we
present the values of the logarithms of $D$ and $\chi_4$ with
\texttt{link2} for observations generated by an $n$-body propagation
and with different astrometric errors. Note that $\log_{10}0.2 \approx
-0.7$.
Here as in most of the next plots, we select the best solution in
terms of $D$ in case of linkages with more than one solution.  The
values of $D$ and $\chi_4$ are correlated which allows us to estimate
the quality of a preliminary orbit based only on the value of
$\chi_4$.  This is important for processing the ITF where we will not
be able to compute $D$ because we will not know the true solution.
Similar results are obtained with a 2-body propagation.

With \texttt{link3} the correlation between $D$ and $\Delta_\star$ is
even stronger (see Figure~\ref{fig:DcritVSNormLK3}), thus allowing us
to estimate in a better way the quality of preliminary orbits using
the value of $\Delta_\star$ only.

\subsection{Time normalisation of the \texorpdfstring{$\chi_4$}{chi4} norm}

\begin{figure}[ht!]
  \centering
  \includegraphics[width=0.66\textwidth]{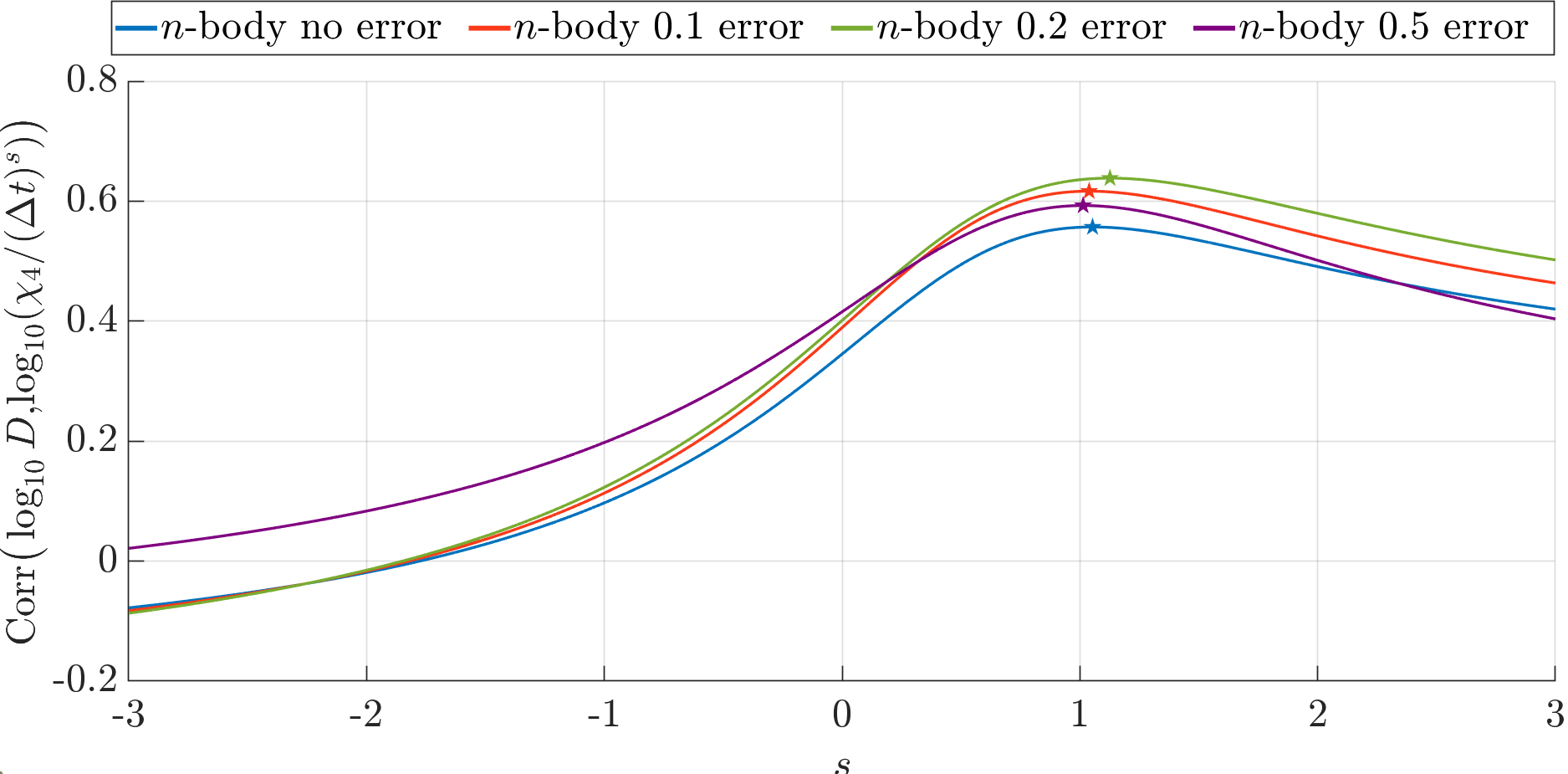}
  \caption{Correlation of the logarithms of $D$ and $\chi_4$/$(\Delta
    t )^s$ as a function of $s$ for \texttt{link2}. The synthetic data
    are generated by an $n$-body propagation with no error, and
    0.1$\arcsec$, 0.2$\arcsec$, 0.5$\arcsec$ error.}
  \label{fig:DcritVSChi4sNorm}
\end{figure}

From the previous analysis, we found that the $\Delta_\star$ norm is
more correlated with the value of $D$ than the $\chi_4$ norm.  Note
that the time separation between the tracklets appears in the
definition of $\Delta_\star$ but not in that of $\chi_4$ (see Sections
\ref{sec:chi4} and \ref{sec:starmetric} for more details).  This is a
relevant difference between the two norms since uncertainty
accumulates over time. To account for this effect we divide $\chi_4$
by the time separation $\Delta t$ between the tracklets raised to
different powers, i.e. $\chi_4/(\Delta t)^s$ with $s\in\mathbb{R}$.
In Figure \ref{fig:DcritVSChi4sNorm} we plot the correlation of the
logarithms of $D$ and $\chi_4/(\Delta t )^s$ as a function of $s$, for
the same data sets used in Figure~\ref{fig:DcritVSNormLK2}.  We see
that to maximise the correlation we can take $s\approx 1$.  Indeed,
the same result still holds for all the synthetic data.  We refer to
the new norm $\chi_4/\Delta t$ as the normalised $\chi_4$.

\begin{figure}[ht!]
  \centering
  \includegraphics[width=0.33\textwidth]{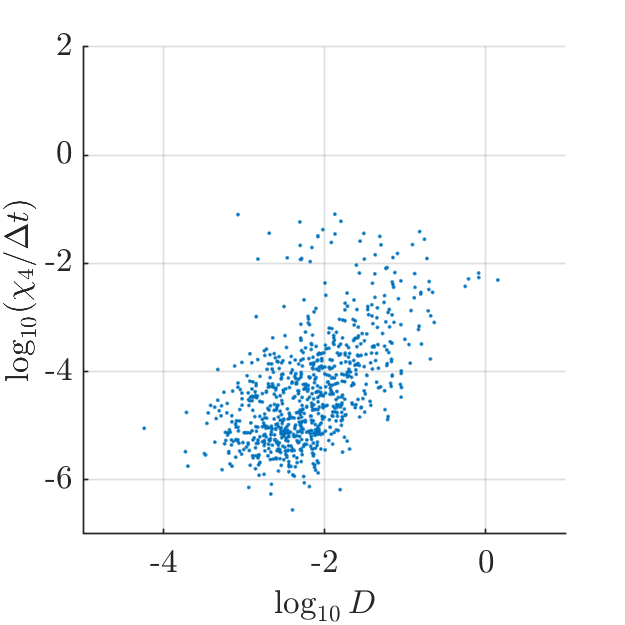}
  \includegraphics[width=0.33\textwidth]{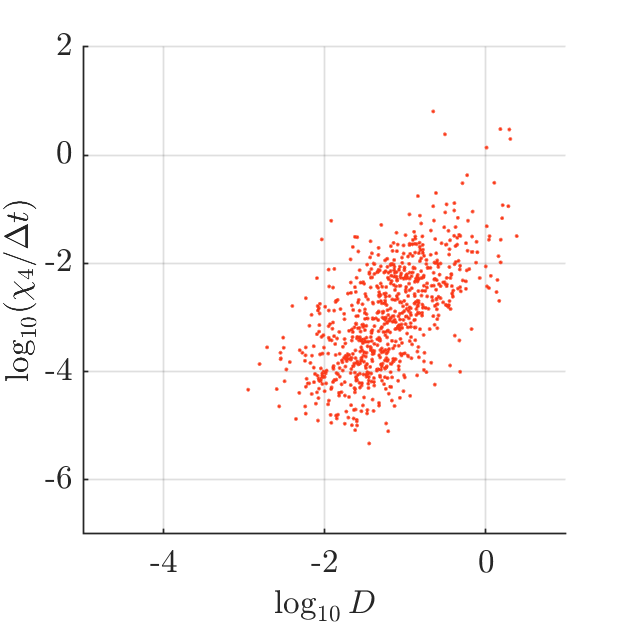}\\
  \includegraphics[width=0.33\textwidth]{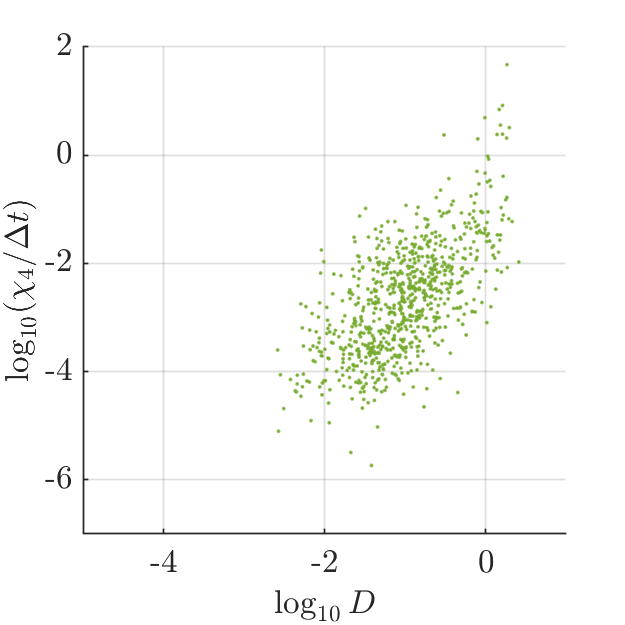}
  \includegraphics[width=0.33\textwidth]{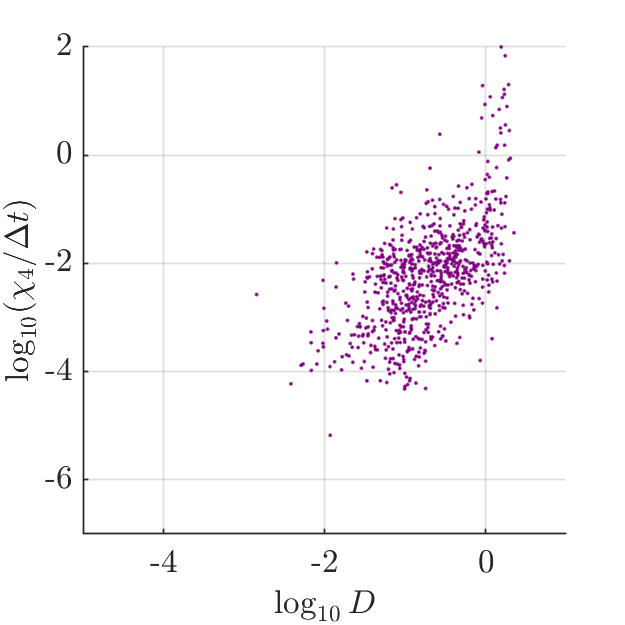}
  \caption{Normalised $\chi_4$ vs $D$ in a log-log plot for
    \texttt{link2}. The synthetic data are generated by an $n$-body
    propagation with no error, and 0.1$\arcsec$, 0.2$\arcsec$,
    0.5$\arcsec$ error (from left to right and top to bottom).}
  \label{fig:DcritVSChiNormNormalized}
\end{figure}

The normalisation of $\chi_4$ by the time separation between the
tracklets produces an increase of the linear relation with the
logarithm of $D$ (compare Figure~\ref{fig:DcritVSChiNormNormalized}
with Figure \ref{fig:DcritVSNormLK2}).

\subsection{The \emph{rms} of the orbit}

Another indicator that can be used to estimate \emph{a priori} the
quality of a preliminary orbit is its \emph{rms} (see
Section~\ref{sec:rms}) with a 2-body propagation of the orbit.

\begin{figure}[ht!]
  \centering
  \includegraphics[width=0.33\textwidth]{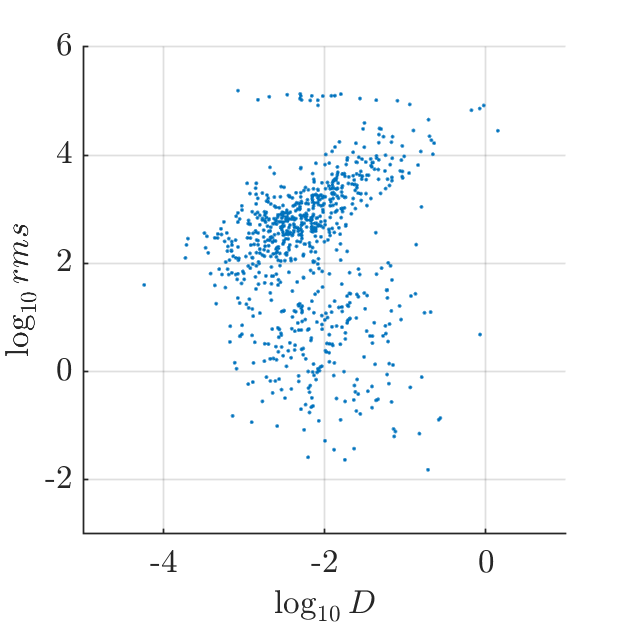}
  \includegraphics[width=0.33\textwidth]{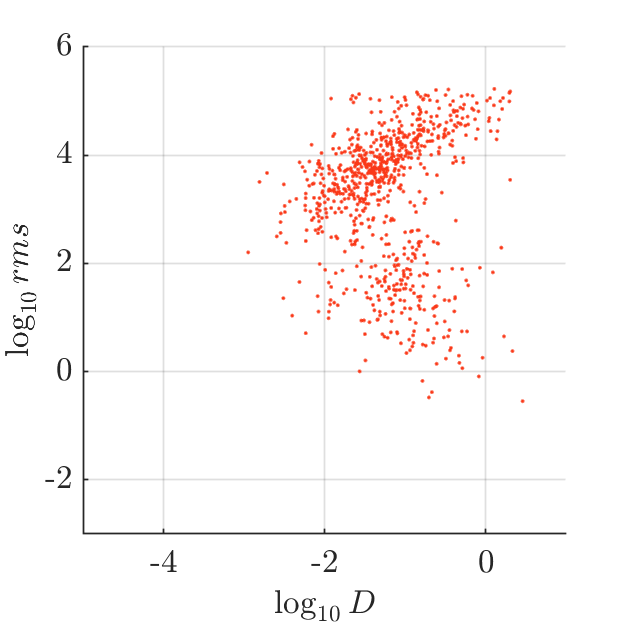}\\
  \includegraphics[width=0.33\textwidth]{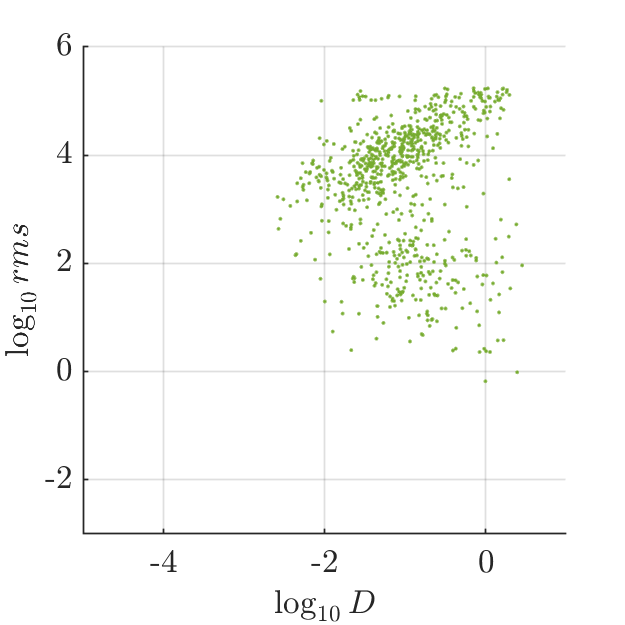}
  \includegraphics[width=0.33\textwidth]{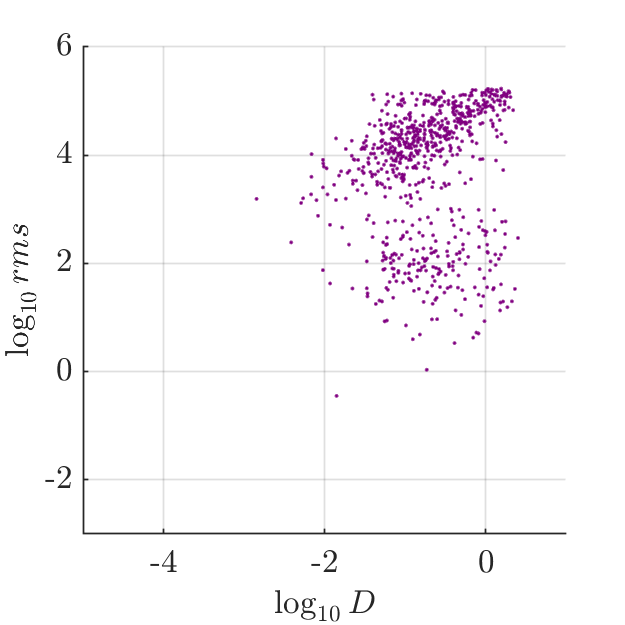}
  \caption{$rms$ vs $D$ in a log-log plot for \texttt{link2}. The
    synthetic data are generated by an $n$-body propagation with no
    error, and 0.1$\arcsec$, 0.2$\arcsec$, 0.5$\arcsec$ error (from
    left to right and top to bottom).}
  \label{fig:rms-true}
\end{figure}

\begin{figure}[ht!]
  \centering
  \includegraphics[width=0.66\textwidth]{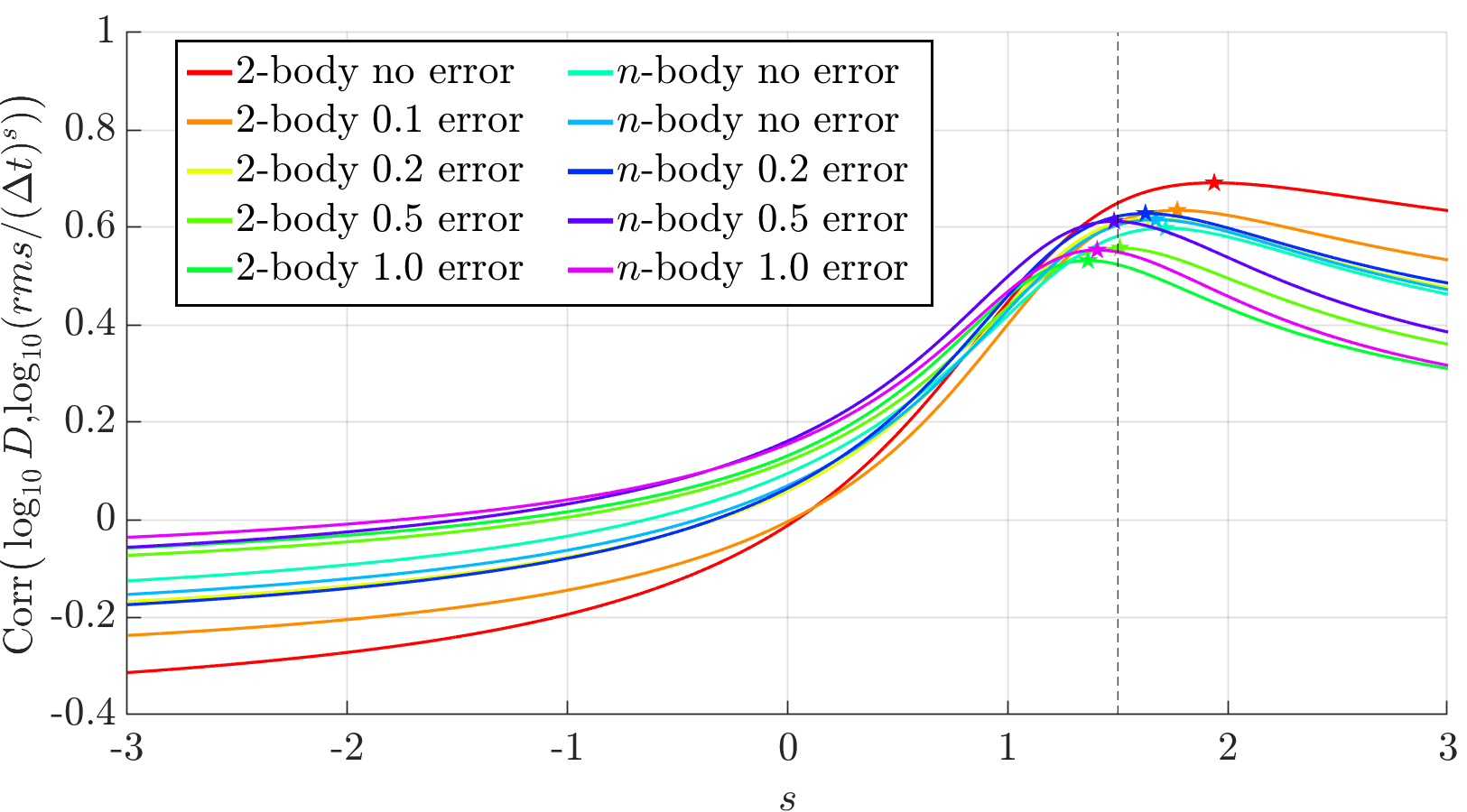}
  \caption{Correlation of the logarithms of $D$ and $rms$/$(\Delta t
    )^s$ in terms of $s$ for \texttt{link2} applied to all the
    synthetic data.}
  \label{fig:bestrmsnormalized}
\end{figure}

Figure \ref{fig:rms-true} shows the values of the logarithms of
\emph{rms} and $D$ with \texttt{link2} for the same data sets as in
Figure~\ref{fig:DcritVSNormLK2}. If we compare the results of this
figure with the similar plots for the normalised $\chi_4$ (see Figure
\ref{fig:DcritVSChiNormNormalized}) and $\Delta_\star$ (see
Figure~\ref{fig:DcritVSNormLK3}), we observe that the correlation
between the logarithms of \emph{rms} and $D$ is lower. This is again
due to the fact that we do not take into account the time between the
tracklets in the computation of the \emph{rms}.  In order to consider
this effect, we follow the same strategy used for the normalisation of
$\chi_4$: the $rms$ is divided by the time separation $\Delta t$
between the tracklets raised to different powers, i.e. $rms/(\Delta
t)^s$ with $s\in\mathbb{R}$.  From Figure \ref{fig:bestrmsnormalized}
it turns out that the best value of this exponent is $s\approx 3/2$
for all the synthetic data, even if the choice is not so evident as
for the $\chi_4$ norm.  We will refer to the new norm $rms/(\Delta
t)^{3/2}$ as to the normalised $rms$.

\begin{figure}[ht!]
  \centering
  \includegraphics[width=0.33\textwidth]{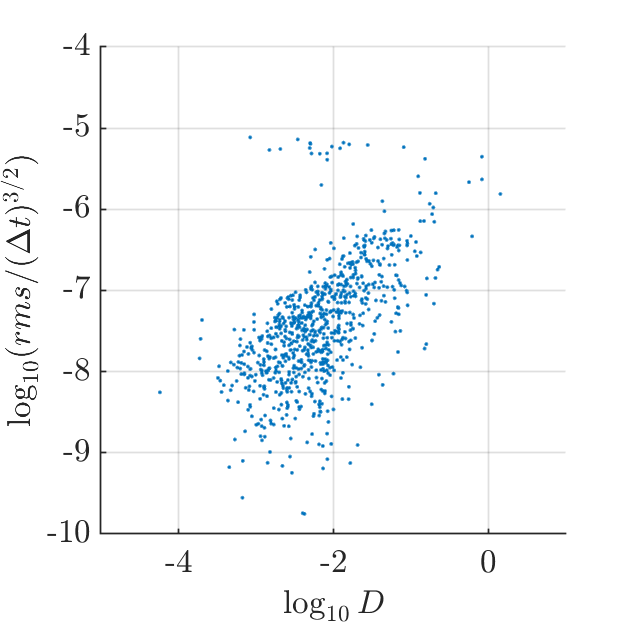}
  \includegraphics[width=0.33\textwidth]{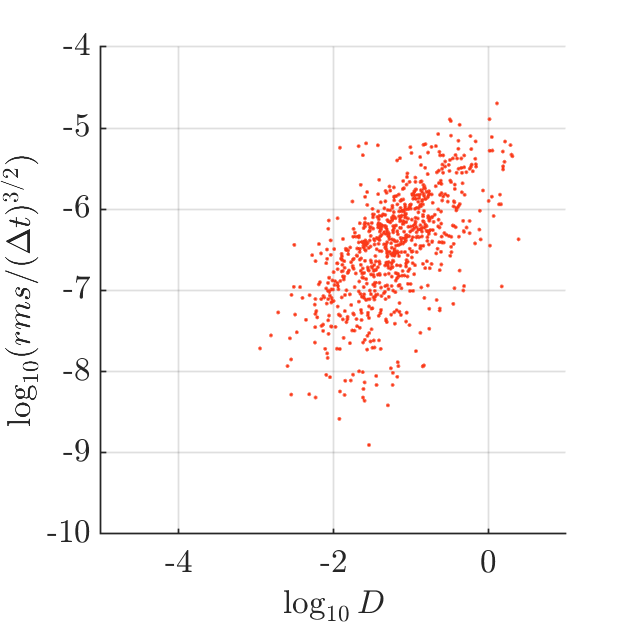}\\
  \includegraphics[width=0.33\textwidth]{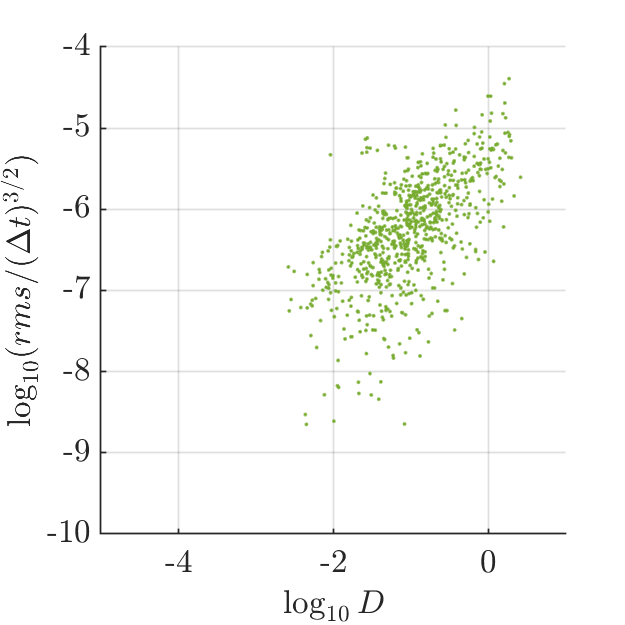}
  \includegraphics[width=0.33\textwidth]{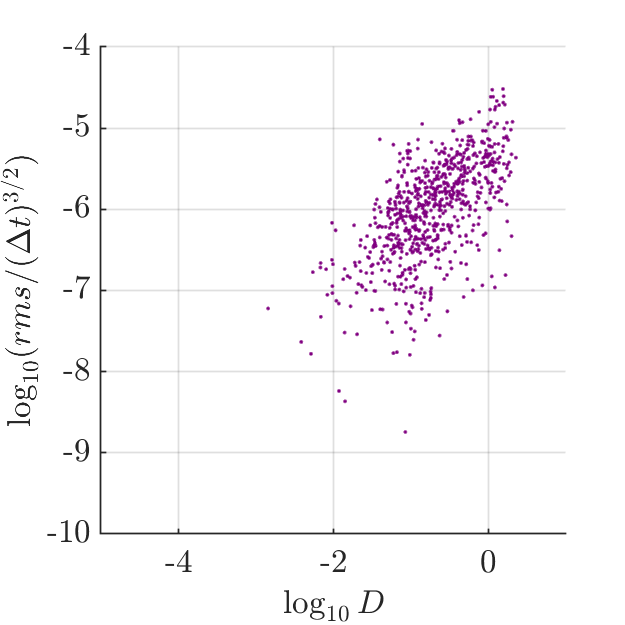}
  \caption{Normalized $rms$ vs $D$ in a log-log plot for
    \texttt{link2}. The synthetic data are generated by an $n$-body
    propagation with no error, and 0.1$\arcsec$, 0.2$\arcsec$,
    0.5$\arcsec$ error (from left to right and top to bottom).}
  \label{fig:rms-true-normalized}
\end{figure}

In Figure~\ref{fig:rms-true-normalized} we present the values of the
logarithms of the normalised $rms$ and $D$ with \texttt{link2}.
Comparison with Figure~\ref{fig:rms-true} shows a better correlation
when time normalisation is applied.

\subsection{Differential corrections of the preliminary orbits}

The previous analysis shows that \texttt{link2} provides better
solutions than \texttt{link3}, but an important aspect which has not
been explored is whether these preliminary solutions are good enough
to obtain least squares orbits, i.e. whether they allow the
differential corrections to converge.

\begin{figure}[ht!]
  \centering
  \includegraphics[width=0.33\textwidth]{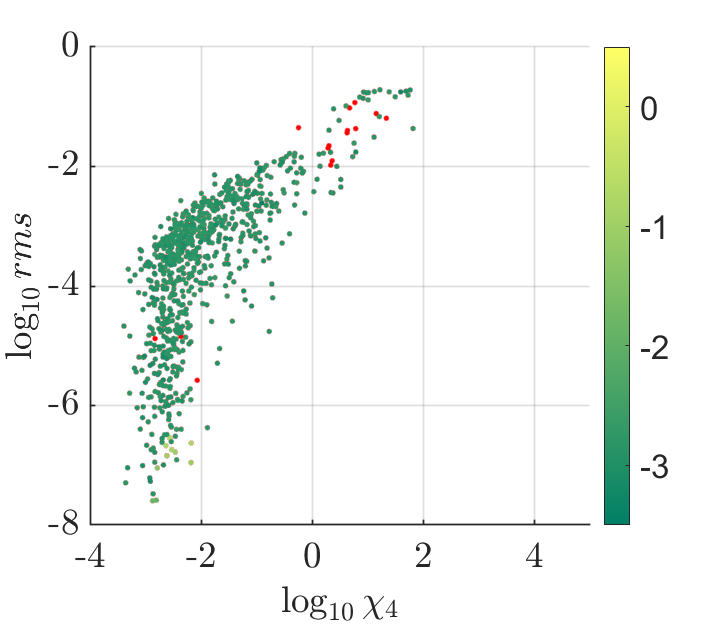}
  \includegraphics[width=0.33\textwidth]{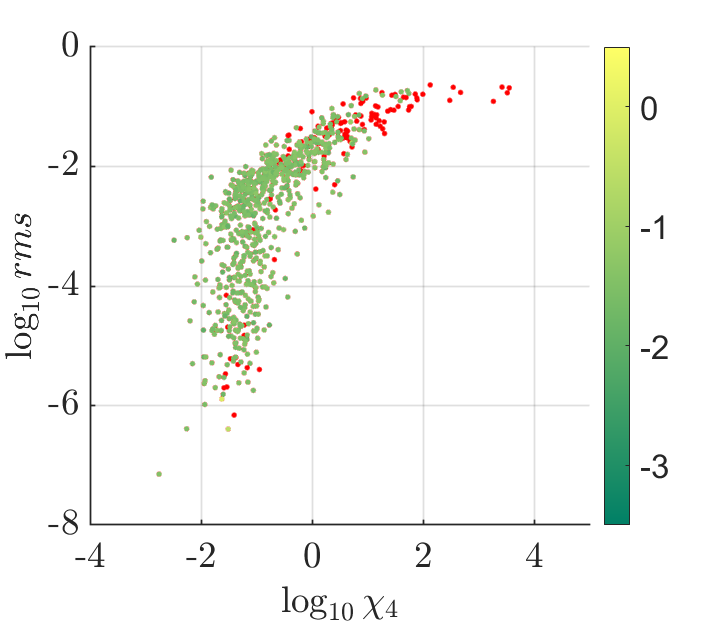}\\
  \includegraphics[width=0.33\textwidth]{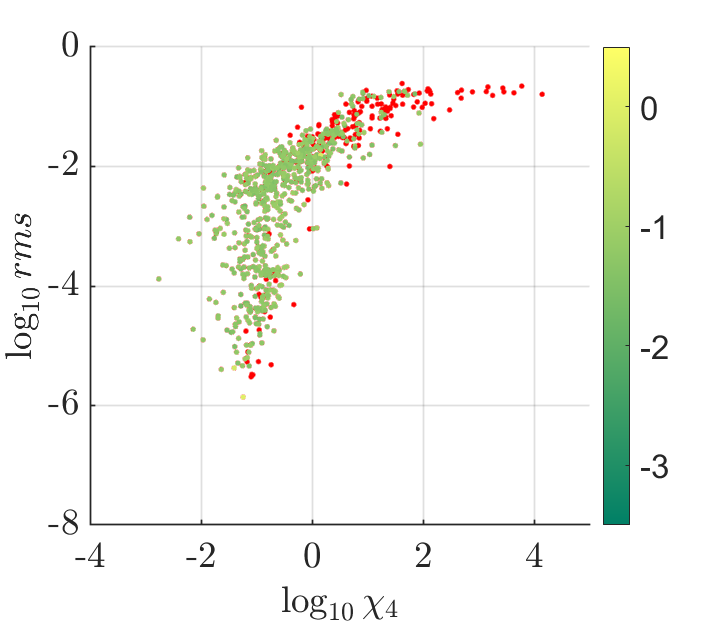}
  \includegraphics[width=0.33\textwidth]{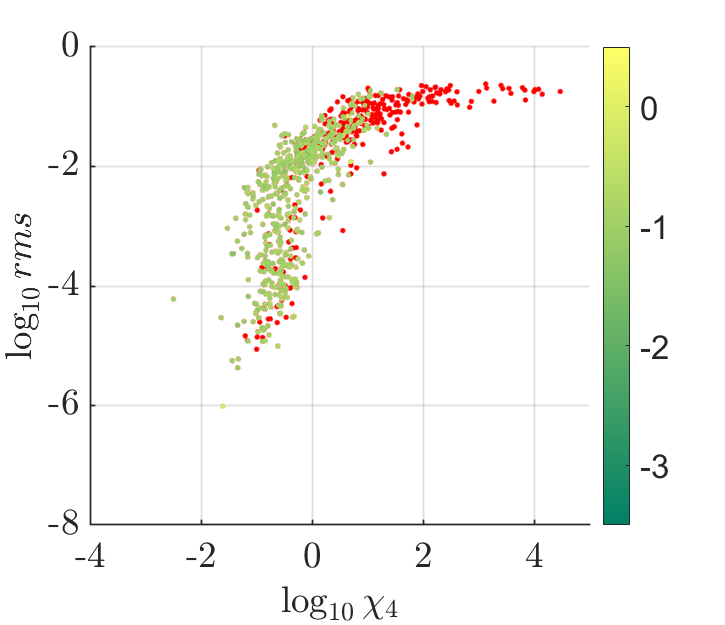}
  \caption{$rms$ vs $\chi_4$ in a log-log plot for the solutions of
    \texttt{link2} and logarithmic values of $R_{LS}$ (green scale) of
    the least squares orbits obtained from these solutions.  The
    synthetic data are generated by an $n$-body propagation with no
    error, and 0.1$\arcsec$, 0.2$\arcsec$, 0.5$\arcsec$ error (from
    left to right and top to bottom).  The red dots correspond to
    preliminary orbits that do not converge in the differential
    corrections scheme.}
  \label{fig:difCorTruenbody}
\end{figure}

We computed a least squares orbit for the best orbit in terms of $D$
obtained with \texttt{link2}.  More than 97\% of the least squares
orbits converged when the data had no astrometric error but the
percentage decreases as the error increases. The values of the least
squares norm $R_{LS}$ (see Section~\ref{sec:LS-rms}) are small using
synthetic data with no error and increase with the error as expected
(see Figure~\ref{fig:difCorTruenbody}).

\subsection{Selection of true linkages}
\label{sec:TLinks}

When using the KI methods in practice it can not be known whether
tracklets belong to the same object and it is desirable to maximise
the number of true linkages while discarding as many of the false ones
as possible.  For this purpose, we try to link \emph{all} the pairs
(for \texttt{link2}) and triplets (for \texttt{link3}) of tracklets
obtainable from those produced for the 822 MBAs.

\begin{figure}[ht!]
  \centering
  \includegraphics[width=0.33\textwidth]{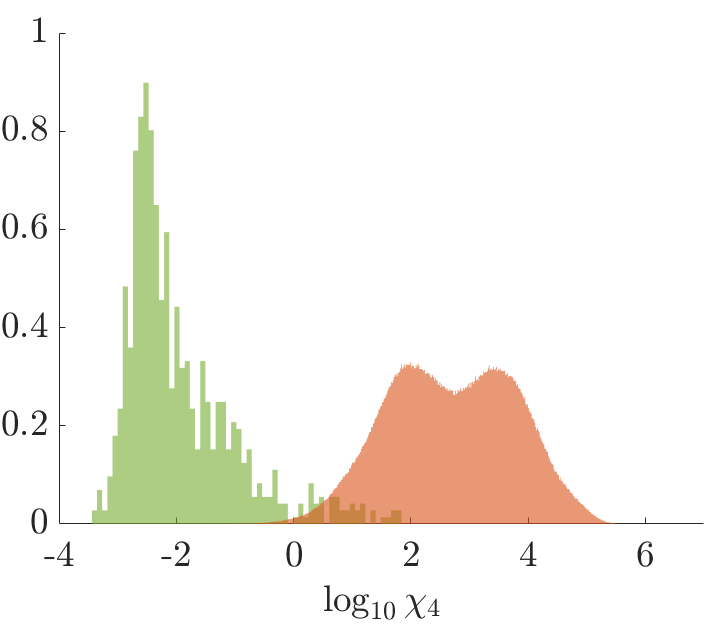}
  \includegraphics[width=0.33\textwidth]{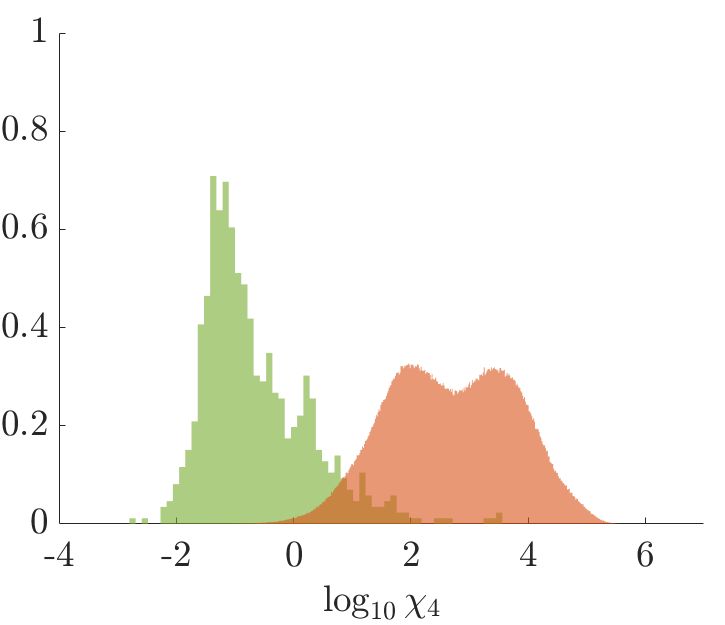}\\
  \includegraphics[width=0.33\textwidth]{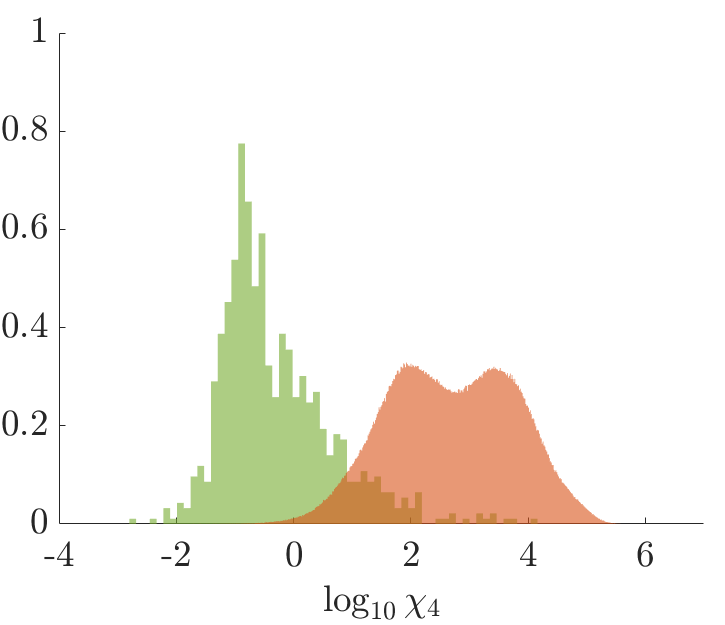}
  \includegraphics[width=0.33\textwidth]{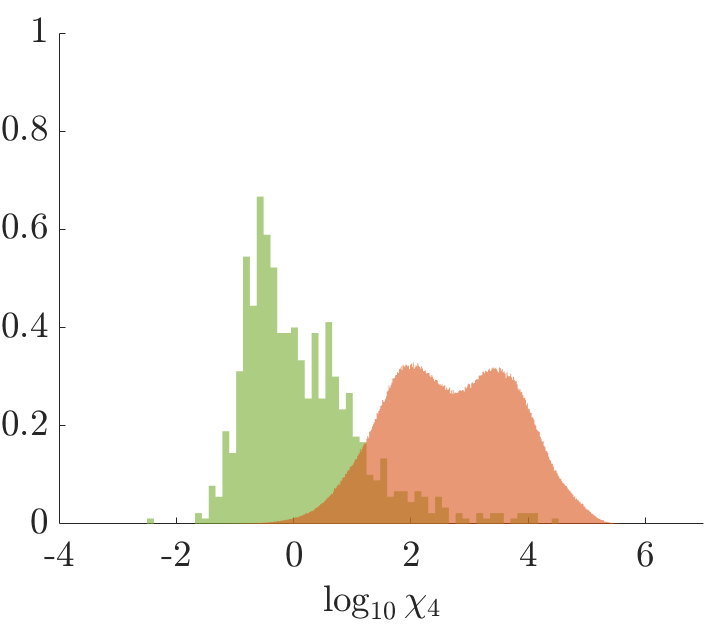}
  \caption{PDF of the logarithm of $\chi_4$ for the true (green) and
    false (red) linkages of \texttt{link2}. The synthetic data are
    generated by an $n$-body propagation with no error, and
    0.1$\arcsec$, 0.2$\arcsec$, 0.5$\arcsec$ error (from left to right
    and top to bottom).}
  \label{fig:pdfLK2truevsfalse}
\end{figure}

Figure~\ref{fig:pdfLK2truevsfalse} shows a clear separation in terms
of $\chi_4$ between true and false linkages for observations with no
error, but the separation decreases as the error increases both with
the 2-body and the $n$-body dynamics.

\begin{figure}
  \centering
  \includegraphics[width=0.66\textwidth]{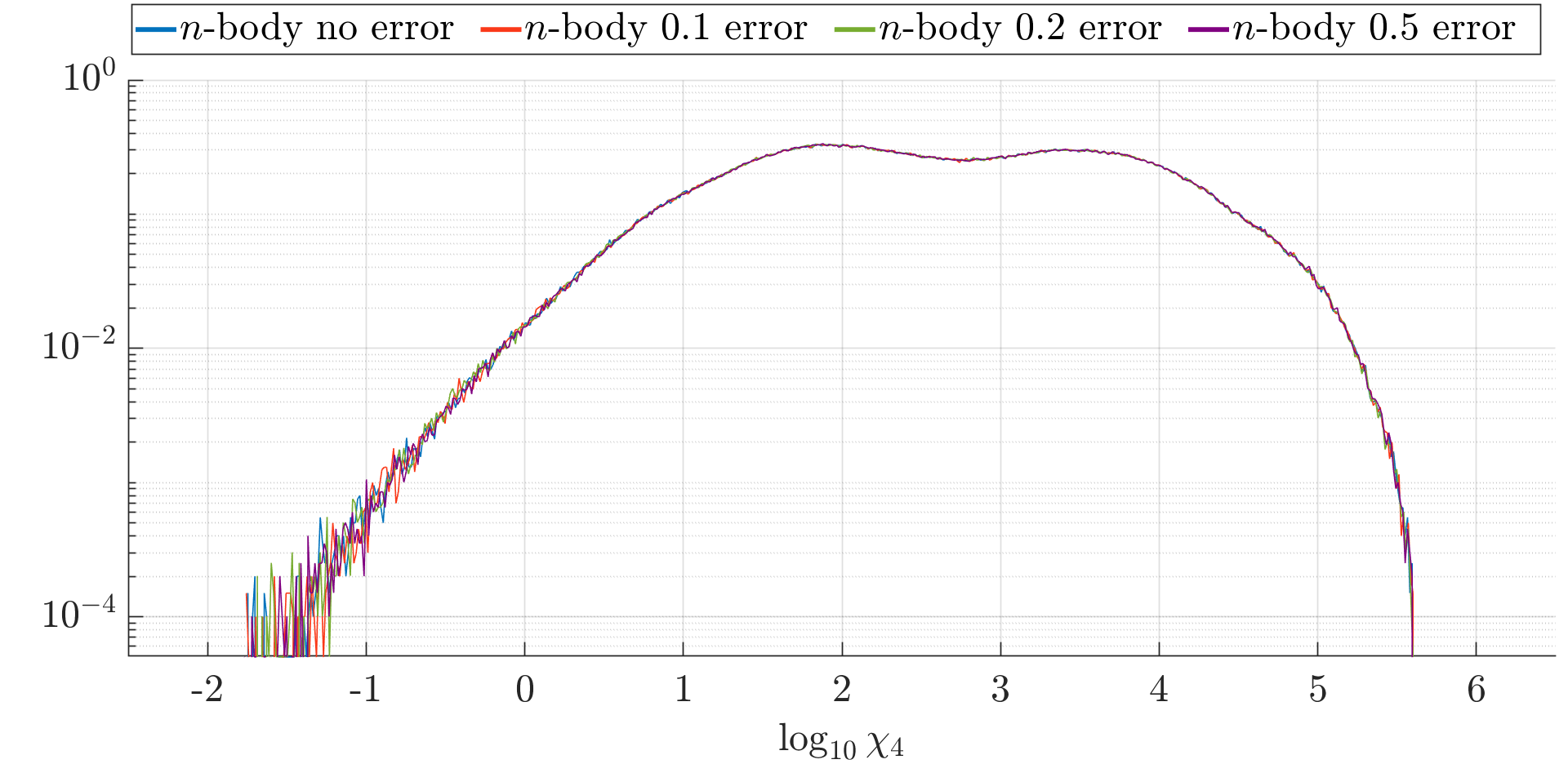}
  \caption{PDF of the logarithm of $\chi_4$ for false linkages using
    \texttt{link2}. The synthetic data are generated by an $n$-body
    propagation with no error and 0.1$\arcsec$, 0.2$\arcsec$,
    0.5$\arcsec$ error.}
  \label{fig:histogramLK2false}
\end{figure}

The selection of a threshold for the $\chi_4$ norm requires care.  For
example, in the case with no error
(Figure~\ref{fig:pdfLK2truevsfalse}) it appears that a suitable value
for $\log_{10}\chi_4$ is 0, however this choice would provide an
overwhelming number of false linkages. It is evident in the PDF of
$\log_{10}\chi_4$ for the false linkages using a logarithmic scale on
the vertical axis (see Figure~\ref{fig:histogramLK2false}). On the
other hand, a smaller threshold, around $-1.4$, yields only a few
outliers, but we expect that even this threshold will generate many
false linkages in large datasets, such as the ITF. In addition, since
the PDF of $\log_{10}\chi_4$ for false linkages remains almost
unchanged as the astrometric error increases, this threshold excludes
more true solutions as the error increases, see Figures
\ref{fig:pdfLK2truevsfalse} and \ref{fig:DcritVSNormLK2}. Of course,
increasing the threshold will generate more false linkages. Similar
conclusions can be drawn for the normalised $\chi_4$, but for the same
number of false linkages we recover less true ones.

\begin{figure}[ht!]
  \centering
  \includegraphics[width=0.33\textwidth]{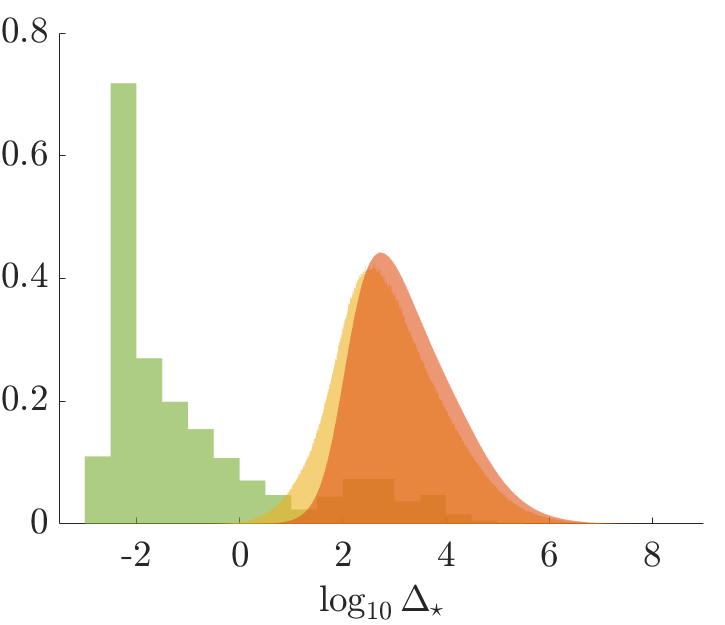}
  \includegraphics[width=0.33\textwidth]{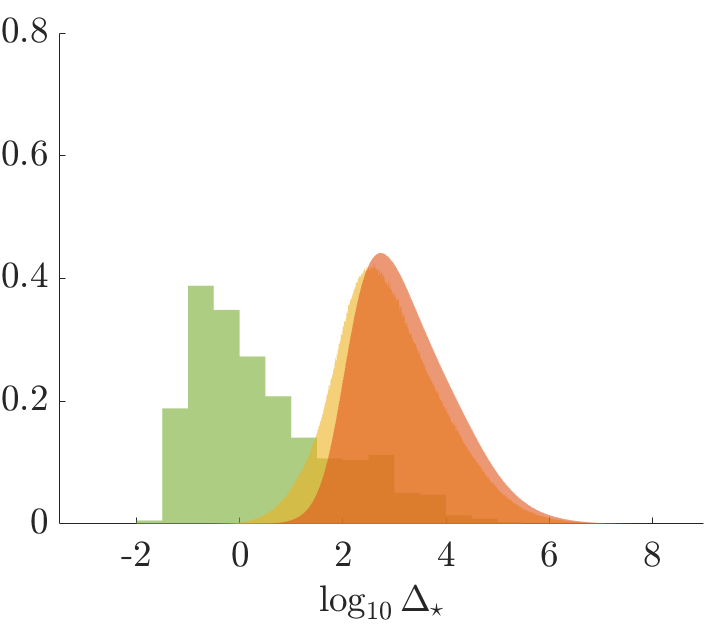}\\
  \includegraphics[width=0.33\textwidth]{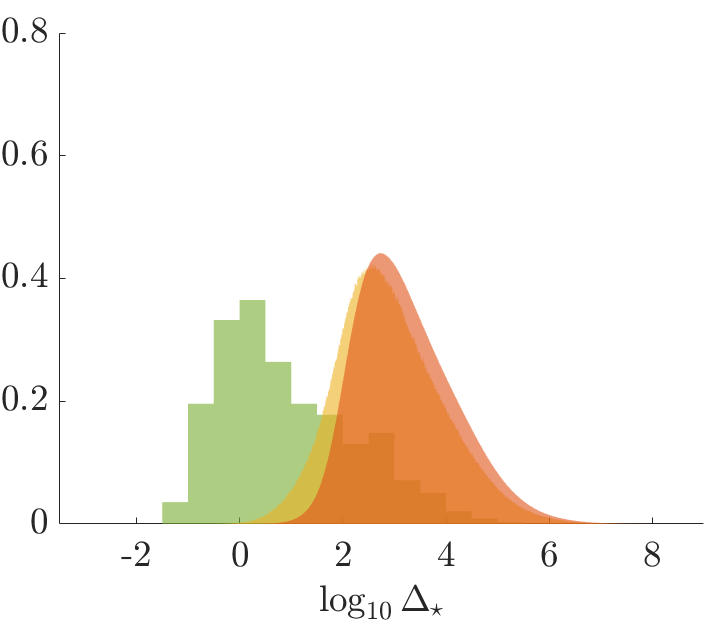}
  \includegraphics[width=0.33\textwidth]{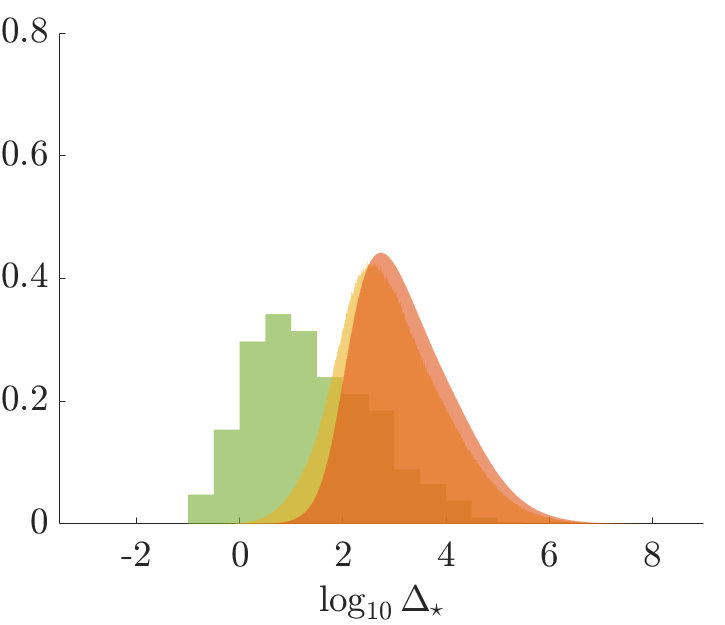}
  \caption{PDF of the logarithm of $\Delta_\star$ for the true (green)
    linkages of \texttt{link3} and the false ones with two tracklets
    belonging to the same object (yellow) and each tracklet belonging
    to different objects (red). The synthetic data are generated by a
    2-body propagation with no error, and 0.1$\arcsec$, 0.2$\arcsec$,
    0.5$\arcsec$ error (from left to right and top to bottom).}
  \label{fig:pdfLK3truevsfalse}
\end{figure}

The results obtained with \texttt{link3} are similar to
\texttt{link2}. In Figure \ref{fig:pdfLK3truevsfalse} we observe a
clear separation in terms of the value of $\Delta_\star$ between the
solutions of the true (green) and false (yellow and red) linkages
considering synthetic data with no astrometric error. The separation
decreases as we increase the error.

\begin{figure}[ht!]
  \centering
  \includegraphics[width=0.66\textwidth]{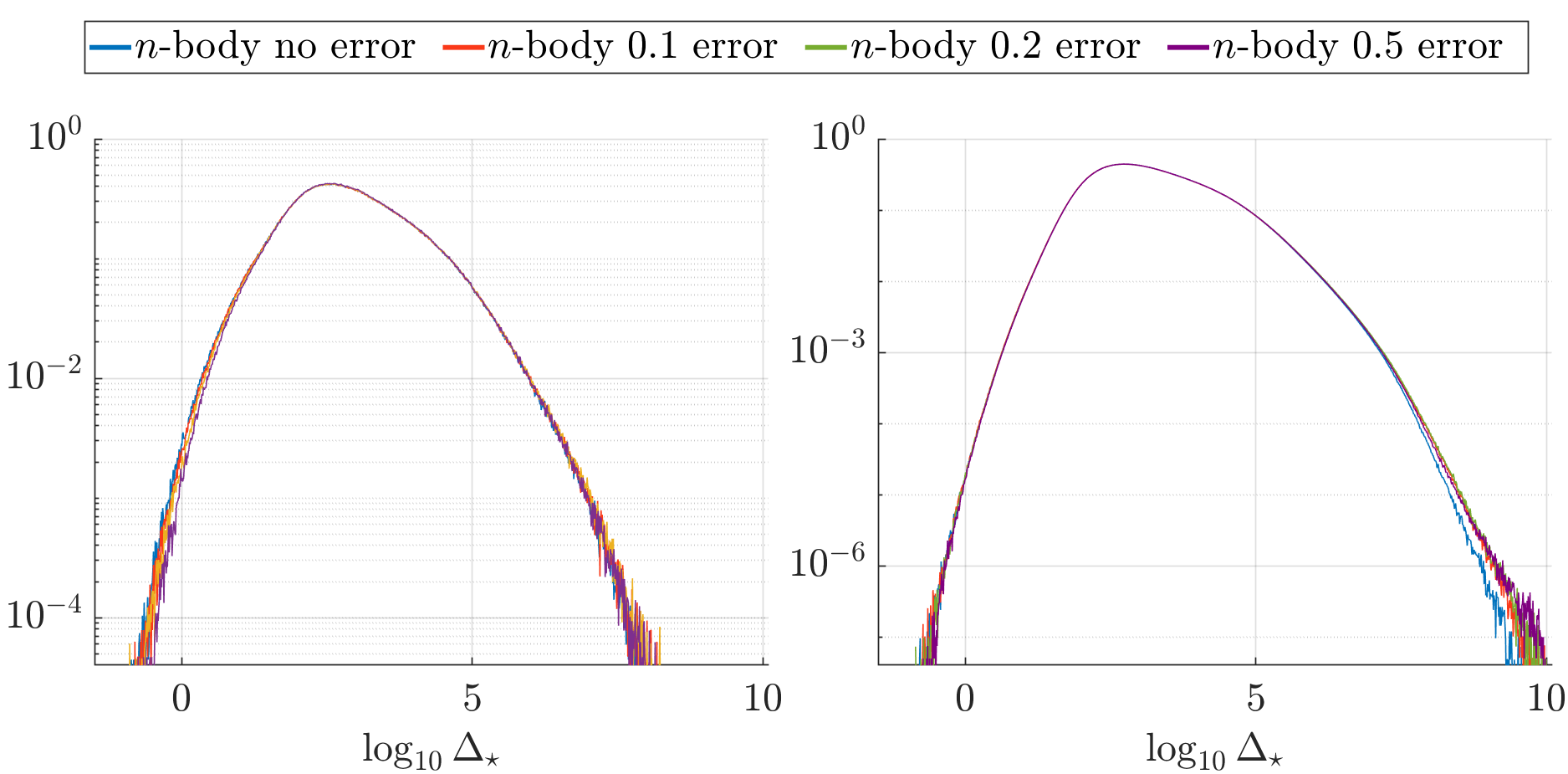}
  \caption{PDF of the logarithm of $\Delta_\star$ for the false
    linkages of \texttt{link3} with two tracklets belonging to the
    same object (left) and each tracklet belonging to different
    objects (right). The synthetic data are generated by a 2-body
    propagation with no error, and 0.1$\arcsec$, 0.2$\arcsec$,
    0.5$\arcsec$ error.}
  \label{fig:histogramLK3false}
\end{figure}

The PDFs of the false solutions are a bit different if the tracklets
belong to two or three different objects (see
Figure~\ref{fig:pdfLK3truevsfalse}), as expected, because the latter
case is ``more wrong'' than the former.  In particular, in Figure
\ref{fig:histogramLK3false} we show in a log-log plot the PDFs of
$\Delta_\star$ for the false linkages corresponding to the yellow
(left) and red (right) distributions of
Figure~\ref{fig:pdfLK3truevsfalse}.

Selecting a threshold around $-0.5$ we have only a few outliers.
However, like for \texttt{link2}, with such a small threshold we lose
a high percentage of true solutions when we include the astrometric
error.

\begin{figure}[ht!]
  \centering
  \includegraphics[width=0.33\textwidth]{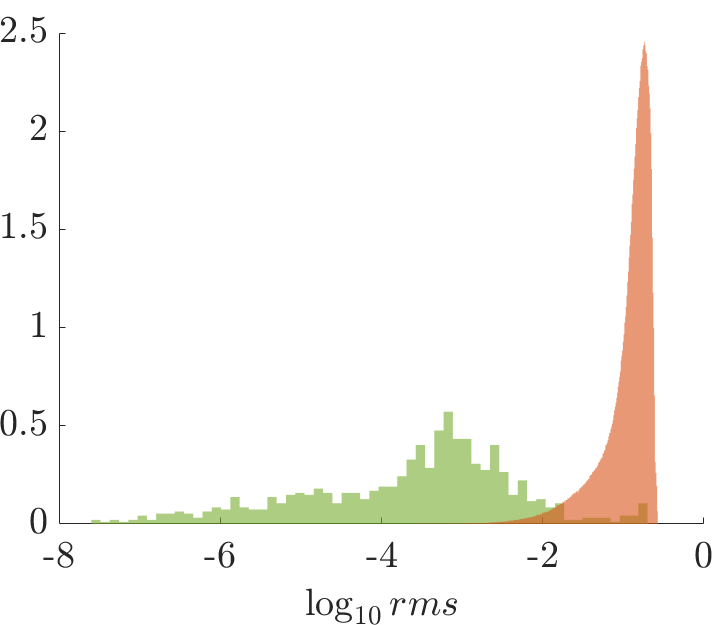}
  \includegraphics[width=0.33\textwidth]{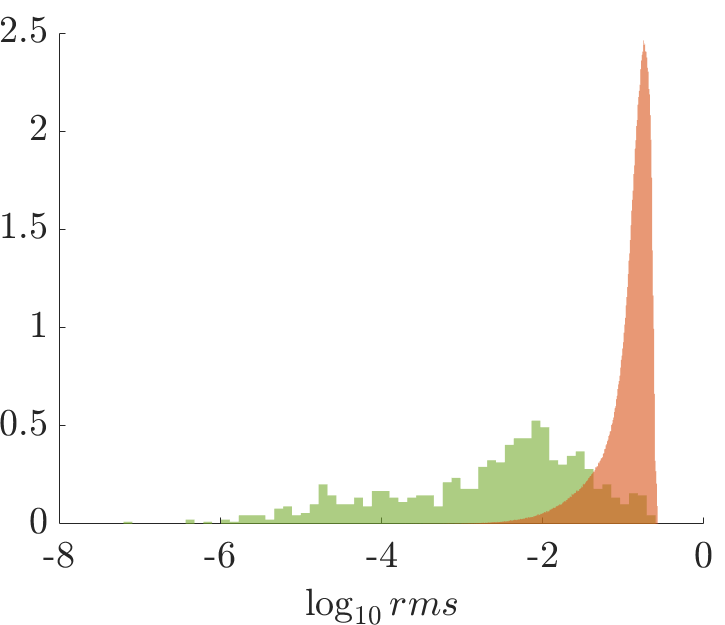}\\
  \includegraphics[width=0.33\textwidth]{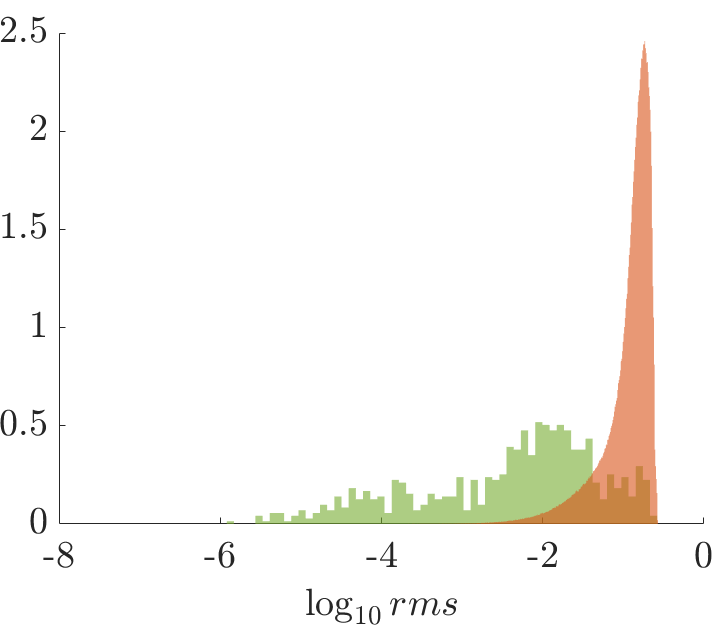}
  \includegraphics[width=0.33\textwidth]{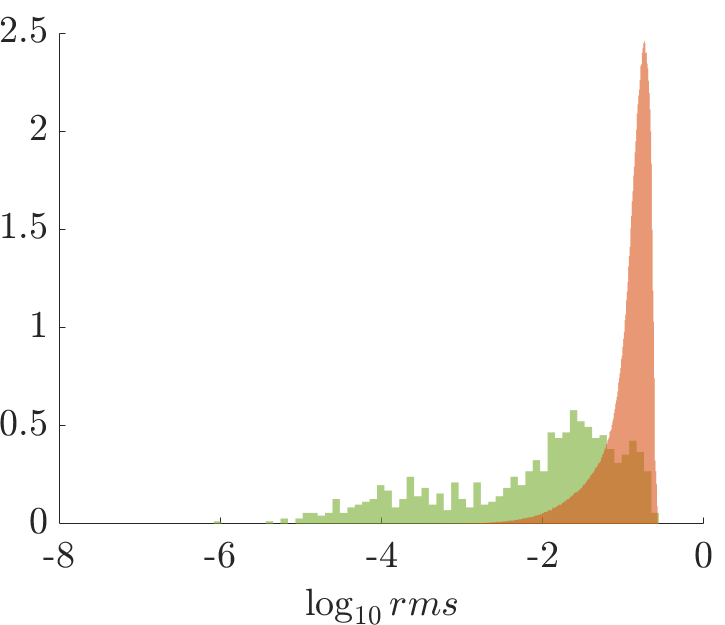}
  \caption{PDF of the logarithm of the $rms$ for the true (green) and
    false (red) linkages of \texttt{link2}. The synthetic data are
    generated by an $n$-body propagation with no error, and
    0.1$\arcsec$, 0.2$\arcsec$, 0.5$\arcsec$ error (from left to right
    and top to bottom).}
  \label{fig:rms-syn}
\end{figure}

Regarding the distribution of the $rms$ of the solutions we have a
similar situation: the separation between the true and false linkages
is clear when we consider data sets with no error and it decreases
with increasing astrometric error (see Figure \ref{fig:rms-syn} for
the results obtained with \texttt{link2}). Moreover, similar to the
case with the normalised $\chi_4$, the normalised \emph{rms} shows a
worse separation between true and false solutions.

\begin{figure}[ht!]
  \centering
  \includegraphics[width=0.33\textwidth]{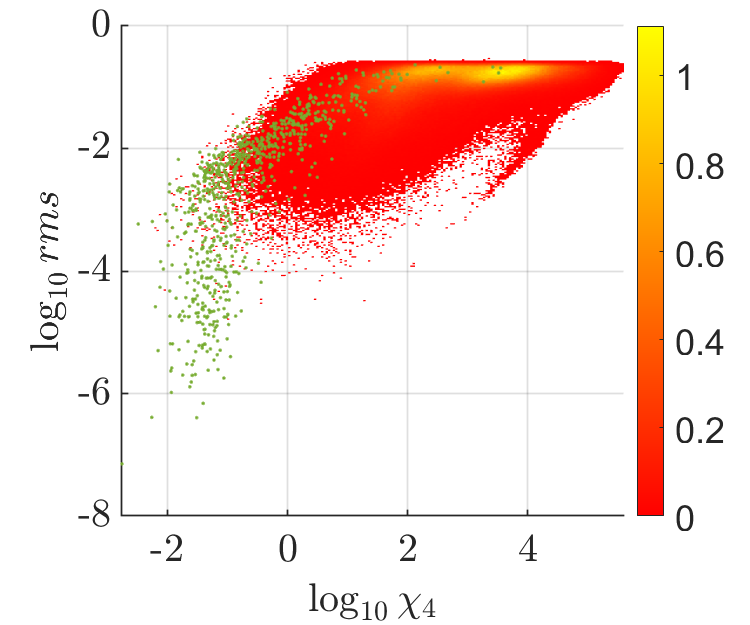}
  \includegraphics[width=0.33\textwidth]{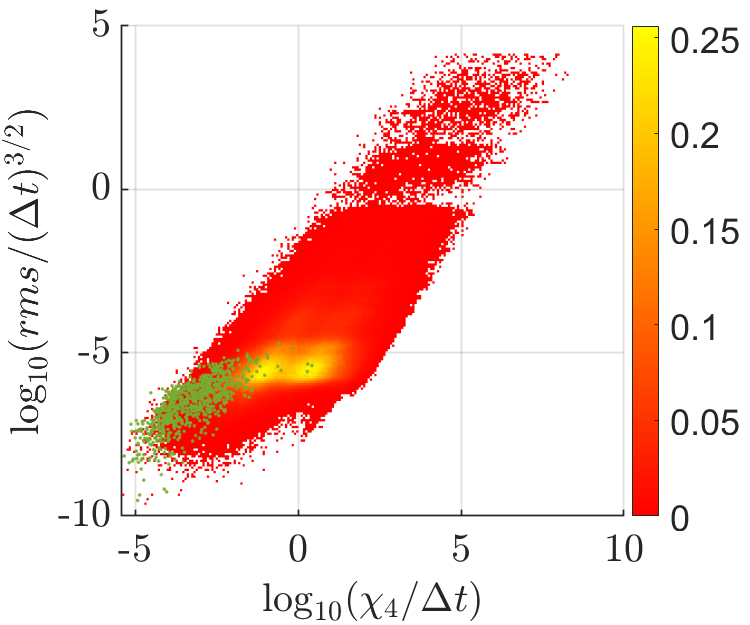}
  \caption{Left: $rms$ vs $\chi_4$ in a log-log plot for the true
    (green) and false (hot colours) linkages of \texttt{link2}. The
    colour scale refers to the values of the PDF of the number of
    false linkages. The synthetic data are generated by an $n$-body
    propagation with 0.1$\arcsec$ error. Right: the same as in the
    left plot but for the normalised $rms$ and normalised $\chi_4$.}
  \label{fig:rms-chi4-syn}
\end{figure}

A combination of all the metrics introduced so far can be used to
discard \emph{a priori} (that is, before applying the differential
corrections scheme) as many false solutions as possible. For example,
in Figure \ref{fig:rms-chi4-syn} the non-normalised (left plot) and
normalised (right plot) metrics could be used to discard some false
linkages. In fact, in the left plot there is a more clear separation
between true and false solutions, while in the right one the true
solutions are gathered in a smaller region of the plane.

\begin{figure}[ht!]
  \centering
  \includegraphics[width=0.33\textwidth]{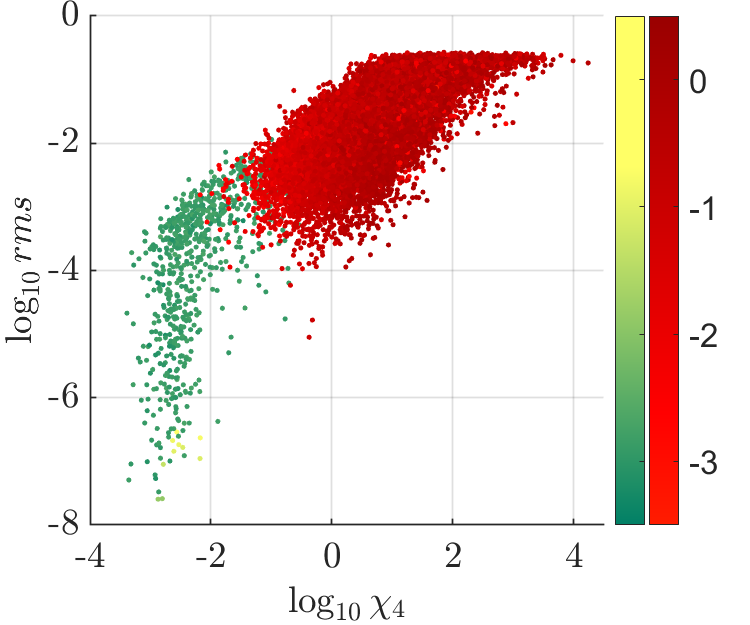}
  \includegraphics[width=0.33\textwidth]{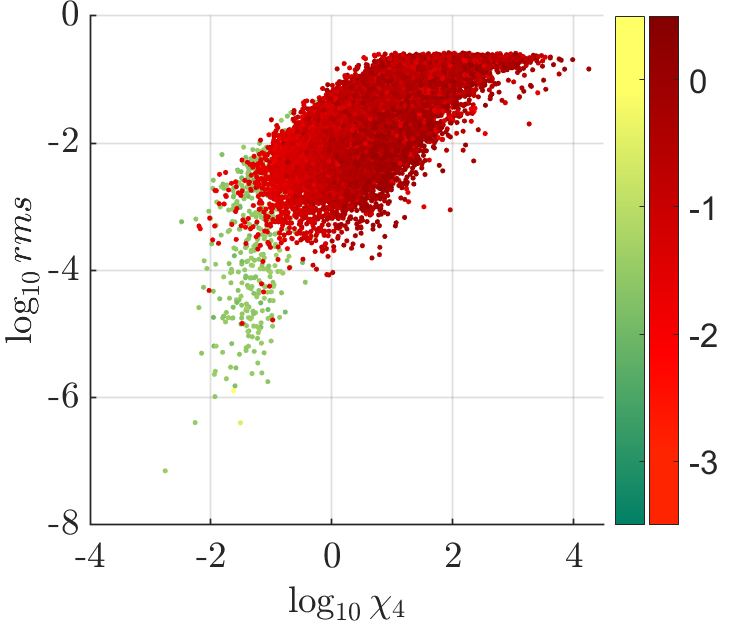}\\
  \includegraphics[width=0.33\textwidth]{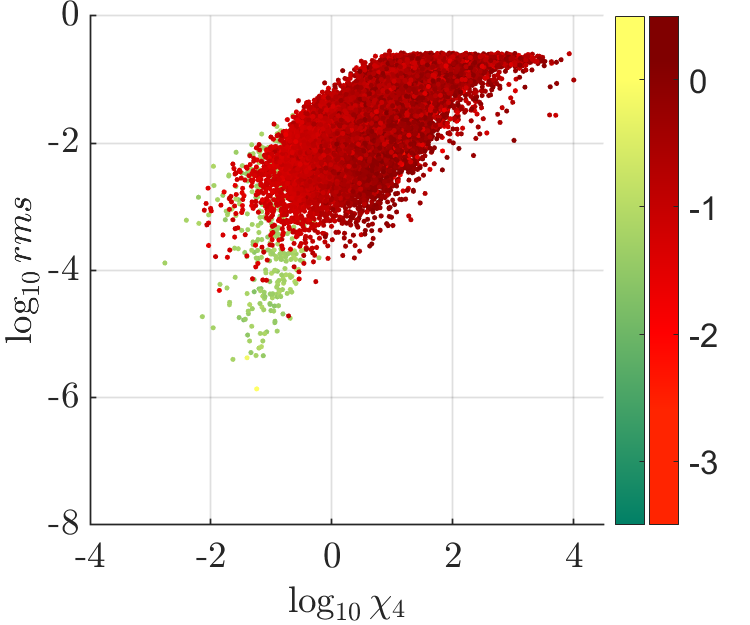}
  \includegraphics[width=0.33\textwidth]{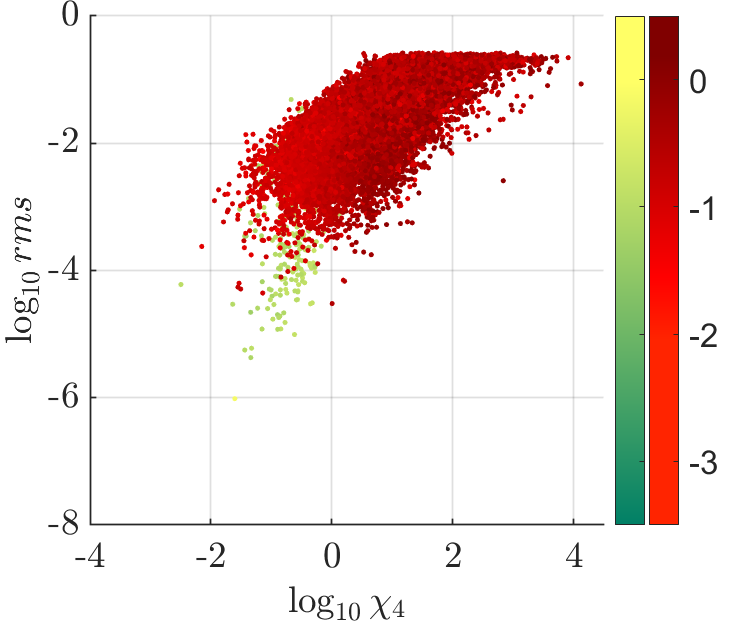}
  \caption{$rms$ vs $\chi_4$ in a log-log plot for the solutions of
    \texttt{link2} and logarithmic values of $R_{LS}$ of the least
    squares orbits obtained from the true (green scale) and false (red
    scale) solutions. The synthetic data are generated by an $n$-body
    propagation with no error, and 0.1$\arcsec$, 0.2$\arcsec$,
    0.5$\arcsec$ error (from left to right and top to bottom).}
  \label{fig:difCorAllnbody}
\end{figure}

Before concluding this section, it is important to note that
discarding false linkages a priori is necessary, because the
combinatorics in large data sets (e.g. the ITF) would require enormous
computing power if all the possible solutions are to be considered
(see section \ref{sec:ITF}). Nevertheless, when differential
corrections are applied to false linkages starting with
\texttt{link2}'s preliminary solutions more than 98\% do not converge.
We also note that all the preliminary solutions with $\log_{10}\chi_4$
greater than 4.3 do not converge. Finally, the separation in the
value of $R_{LS}$ for the solutions obtained from the true and false
linkages is quite clear when there is no error, and it decreases as
the astrometric error increases (see Figure \ref{fig:difCorAllnbody}).

\section{Testing the KI methods with real data}
\label{sec:realdata}

In this section we study the behaviour of \texttt{link2} and
\texttt{link3} using real observations from the Pan-STARRS1 telescope.

\subsection{The error distribution for real observations}
\label{sec:realerror}

Working with real observations we can not assume that the astrometric
errors are independently distributed following a 2d Gaussian
distribution as in the case of the synthetic observations \citep[see
  for example][]{Baer2011,Carpino2003}.

\begin{figure}[ht!]
  \centering
  \includegraphics[width=0.22\textwidth]{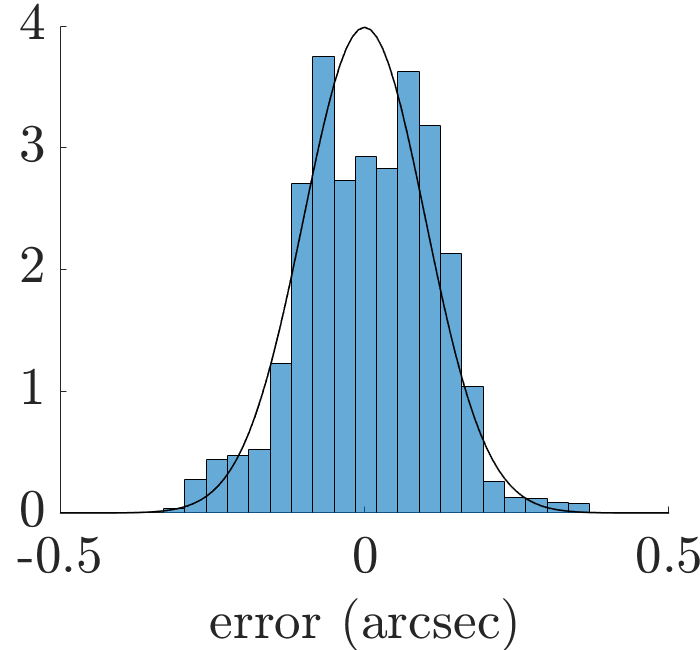}
  \includegraphics[width=0.22\textwidth]{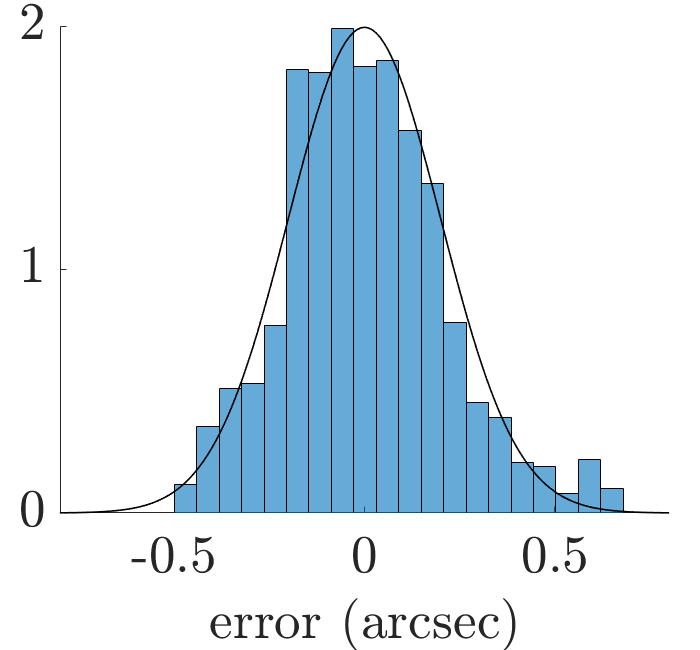}
  \includegraphics[width=0.22\textwidth]{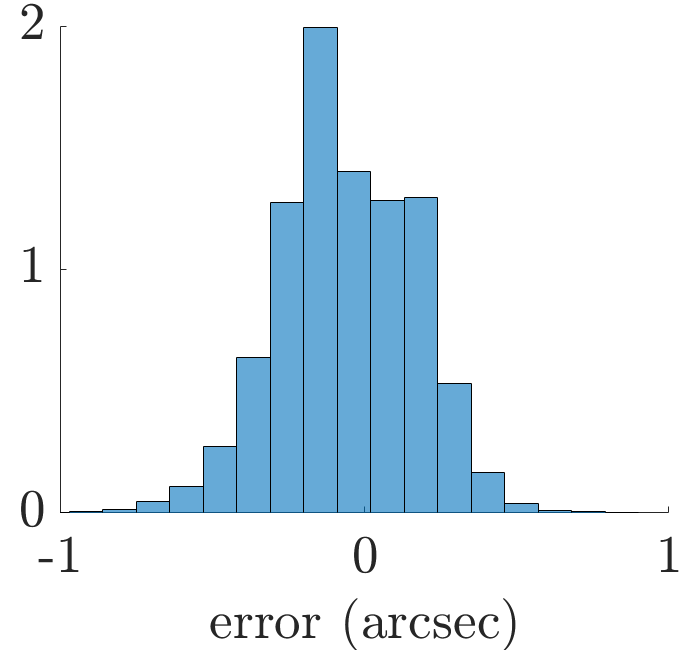}
  \caption{Normalised histograms of the astrometric error for the
    synthetic data generated by an $n$-body propagation with
    0.1$\arcsec$ and 0.2$\arcsec$ error, and for the real data (from
    left to right). The black curve in the left and middle plots
    represents the normal distribution with zero mean and standard
    deviation of 0.1$\arcsec$ and 0.2$\arcsec$, respectively.}
  \label{fig:histogramErrorObs}
\end{figure}

The astrometric error was introduced in the synthetic observations by
applying the following procedure.  First, we simulate perfect
observations $(\alpha_i^*,\delta_i^*)$ using $2$-body or $n$-body
propagation.  Then, we select a random angle, $\theta\in[0,2\pi)$,
  representing a direction on the tangent plane to the celestial
  sphere, and change the observations following a Gaussian
  distribution in that direction.  In this way, the astrometric error
  in the observation $(\alpha_i,\delta_i)$, that we define as
  \begin{equation}\label{eq:astromError}
    e_i = \text{sign}(\Delta_{\alpha_i})\sqrt{\Delta_{\alpha_i}^2\cos^2\delta_i
    + \Delta_{\delta_i}^2},
  \end{equation}
where $\Delta_{\alpha_i} = \alpha_i^*-\alpha_i$ and $\Delta_{\delta_i}
= \delta_i^*-\delta_i$, follows a normal distribution with zero mean
and standard deviation $\sigma$ in the synthetic data (see for example
the left and middle plots in Figure \ref{fig:histogramErrorObs}).

\begin{table}[ht!]
  \centering
  \begin{tabular}{lrrrr}
    Dataset & \multicolumn{1}{l}{Expec. $\mu$} & \multicolumn{1}{l}{Compu. $\mu$}&
    \multicolumn{1}{l}{Expec. $\sigma$} & \multicolumn{1}{l}{Compu. $\sigma$} \\ \hline
    $n$-body 0.1 & 0 & 0.0073       & 0.1  & 0.1088        \\
    $n$-body 0.2 & 0 & 0.0037       & 0.2  & 0.2099        \\
    real         & - & 0.0593       & -    & 0.2227        \\
  \end{tabular}
  \caption{Statistics of the distribution of the astrometric error
    (values are in arcsec).}
  \label{tab:errorObs}
\end{table}

The error in the real observations is estimated from equation
\eqref{eq:astromError}, where the perfect observations
$(\alpha_i^*,\delta_i^*)$ are calculated by a full $n$-body
propagation \citep{Granvik2009-OpenOrb} which also includes the effect
of the largest asteroids.  The resulting distribution of the errors
looks like a Gaussian (see the right plot of Figure
\ref{fig:histogramErrorObs}), and its mean and standard deviation are
given in Table \ref{tab:errorObs}. However, the important difference
with respect to the synthetic data is the correlation of the errors in
observations within the same tracklet which is no longer negligible.
Indeed, in the case of real data there is a significant correlation
($>0.5$) between the errors of two observations randomly selected from
the same tracklet (see Table \ref{tab:covarianceObs}).  This means
that the values of the indicators ($\chi_4$, $\Delta_\star$,
\emph{rms}) will be distributed differently for real observations
compared to synthetic ones, since we also assume that the astrometric
error is uncorrelated in the construction of the attributables.

\begin{table}[ht!]
  \centering
  \begin{tabular}{lrrr}
    & \multicolumn{1}{l}{$n$-body 0.1$\arcsec$} & \multicolumn{1}{l}{$n$-body 0.2$\arcsec$} &
    \multicolumn{1}{l}{\phantom{$n$-body}real}  \\ \hline
    Correlation & 0.0365 & 0.0458 & 0.5206  \\
  \end{tabular}
  \caption{Correlation of the errors in the observations that belong
    to the same tracklet.}
  \label{tab:covarianceObs}
\end{table}

\begin{figure}[ht!]
  \centering
  \includegraphics[width=0.66\textwidth]{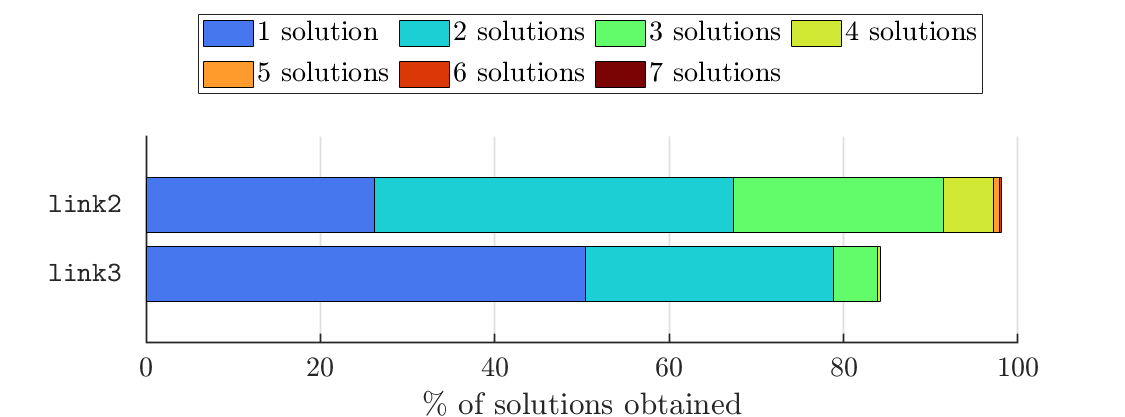}
  \caption{Percentage of linkages and multiplicity of the solutions
    obtained by \texttt{link2} and \texttt{link3} with the real data
    set.}
  \label{fig:lk2vslk3solutionsReal}
\end{figure}

\begin{figure}[ht!]
  \centering
  \includegraphics[width=0.19\textwidth]{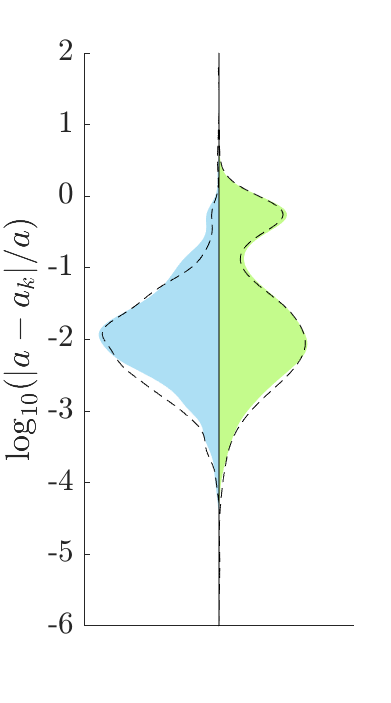}
  \includegraphics[width=0.19\textwidth]{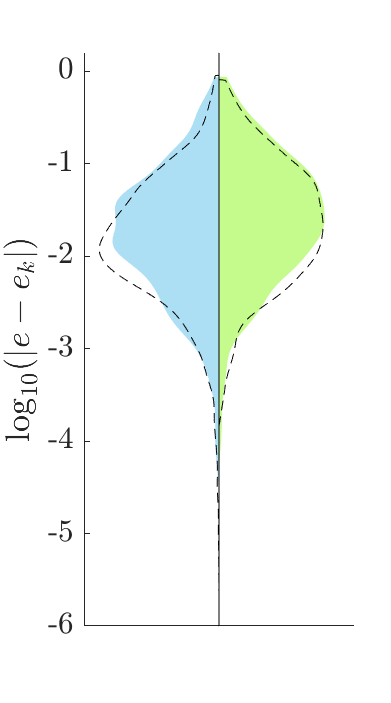}
  \includegraphics[width=0.19\textwidth]{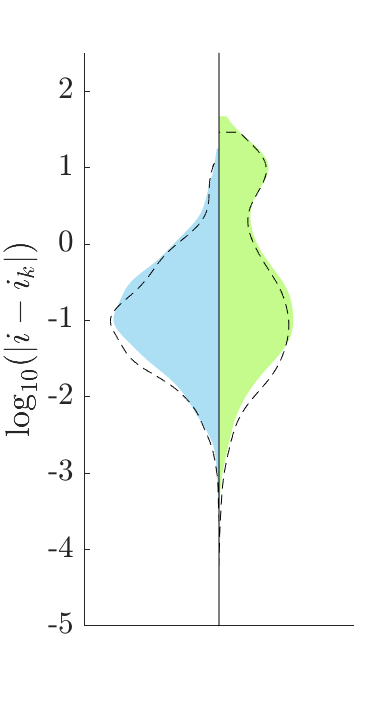}
  \includegraphics[width=0.19\textwidth]{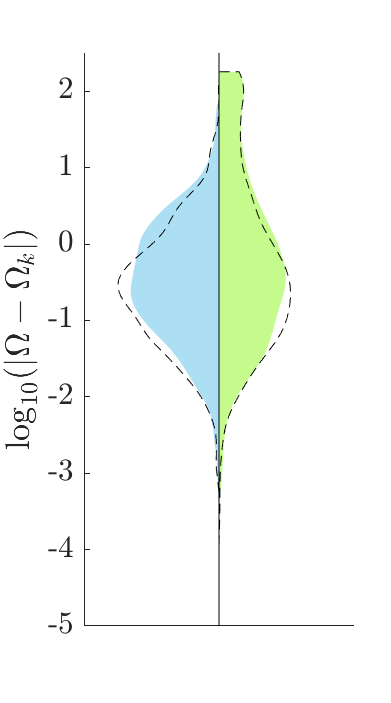}
  \includegraphics[width=0.19\textwidth]{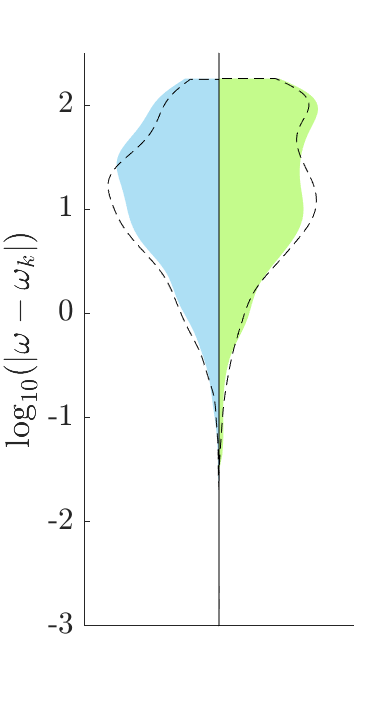}
  \caption{PDF of the logarithm of the error in $a$, $e$, $i$,
    $\Omega$, and $\omega$ (from left to right) for the solutions
    obtained by \texttt{link2} (blue) and \texttt{link3} (green) on
    the real data set. The dashed lines correspond to the PDFs for the
    synthetic data set generated by an $n$-body propagation with
    0.2$\arcsec$ error.}
  \label{fig:lk2vslk3pdfReal}
\end{figure}

\subsection{Numerical results}

We now examine the performance of our \texttt{link2} and
\texttt{link3} algorithms on real data.

\begin{figure}[ht!]
  \centering
  \includegraphics[width=0.33\textwidth]{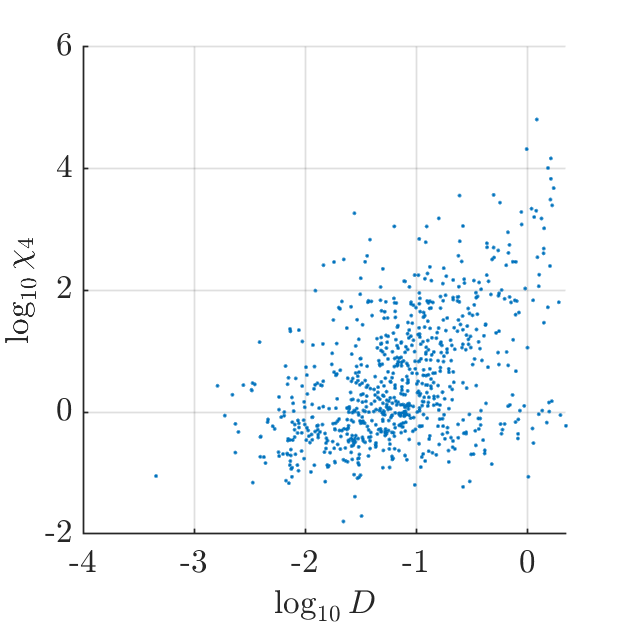}
  \includegraphics[width=0.33\textwidth]{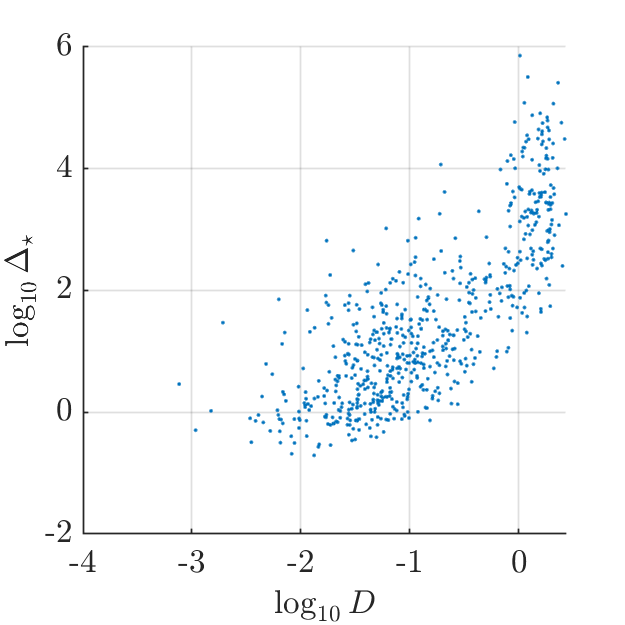}
  \caption{Left: $\chi_4$ vs $D$ in a log-log plot for \texttt{link2}
    with real data. Right: $\Delta_\star$ vs $D$ in a log-log plot
    for \texttt{link3} with real data.}
  \label{fig:DcritVSNormReal}
\end{figure}

The percentage of true linkages recovered by each method
(Figure~\ref{fig:lk2vslk3solutionsReal}) is similar to the results
obtained with synthetic data with an astrometric error of 0.2$\arcsec$
(see Figure~\ref{fig:lk2vslk3solutions}). In addition, the quality of
the solutions (i.e. the error in the orbital elements
$a,e,i,\Omega,\omega$) is similar (see Figure
\ref{fig:lk2vslk3pdfReal}) and reassures us that it is suitable to
employ synthetic data to study the behaviour of both methods.  On the
other hand, the $\chi_4$ and $\Delta_\star$ norms for the real data
are larger than those in the synthetic data case with 0.2$\arcsec$
error (compare Figure \ref{fig:DcritVSNormReal} with Figures
\ref{fig:DcritVSNormLK2} and \ref{fig:DcritVSNormLK3}) because the
errors in the real observations within a tracklet are not independent.

\begin{figure}[ht!]
  \centering
  \includegraphics[width=0.66\textwidth]{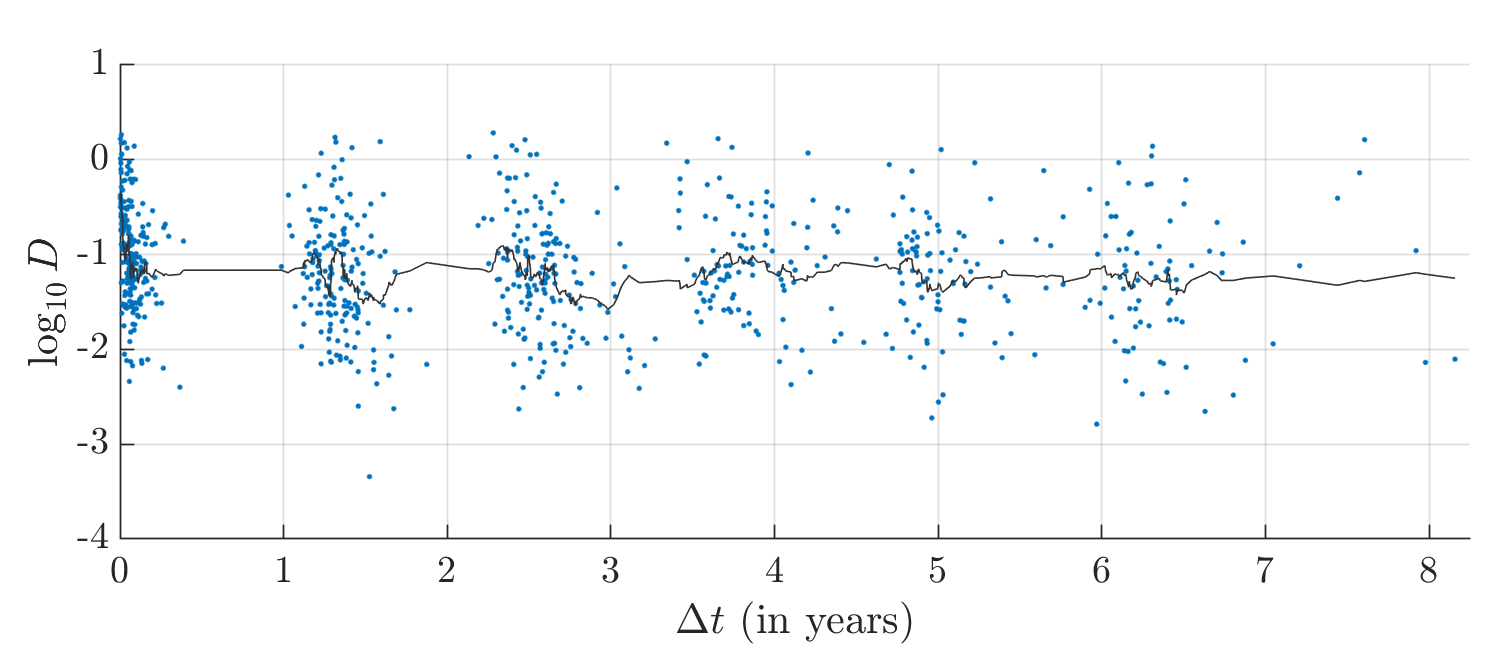}
  \caption{Value of the $D$-criterion of the preliminary orbits
    obtained by \texttt{link2} with the real data as a function of the
    time separation between the tracklets.}
  \label{fig:DvsDeltaT}
\end{figure}

Note that both methods are able to link tracklets with a large time
separation and produce preliminary solutions quite close to the real
orbits even when the tracklets are separated by a few years.
In Figure \ref{fig:DvsDeltaT} we plot the value of the $D$-criterion
of the preliminary orbits obtained with \texttt{link2} in terms of the
time separation between the tracklets. Moreover, we draw a black curve
corresponding to the { \em simple moving average}
$\mathrm{SMA}_{25,12}$, made with 25 data points, 12 on the left and
12 on the right of the central value, when they are available. When
the central value is close to the boundary we use all the available
data points within these limits.  Note that the values of
$\mathrm{SMA}_{25,12}$ keep below the $-0.7$ threshold of the
similarity criterion.

\begin{figure}[ht!]
  \centering
  \includegraphics[width=0.33\textwidth]{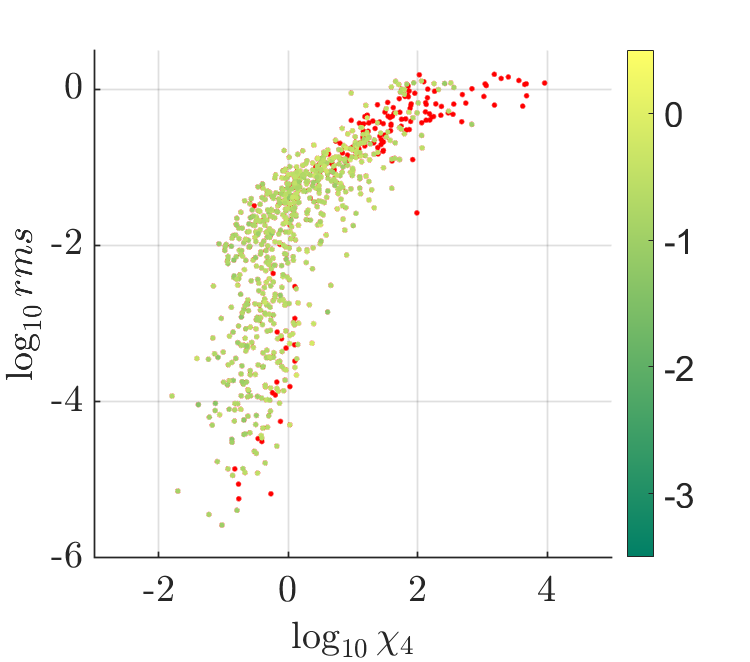}
  \includegraphics[width=0.33\textwidth]{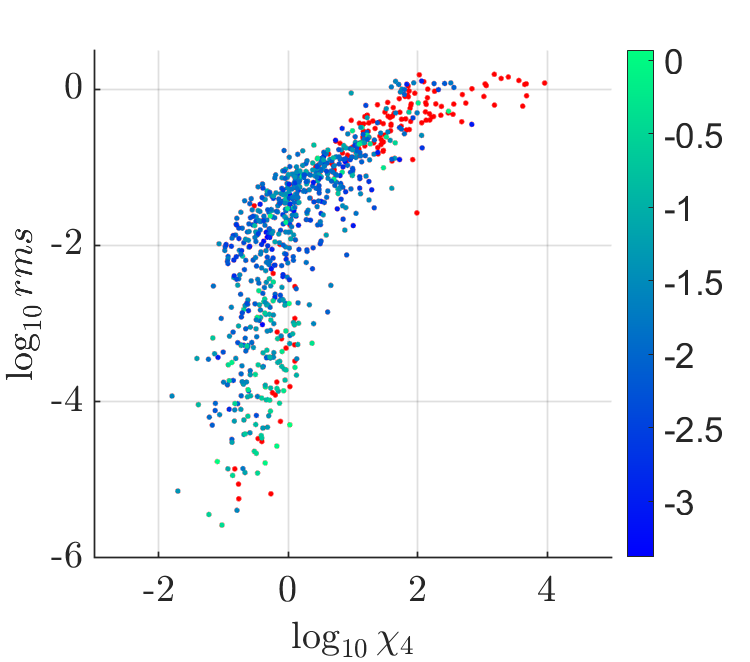}
  \caption{$rms$ vs $\chi_4$ in a log-log plot for the solutions of
    \texttt{link2} using real data and logarithmic values of $R_{LS}$
    (green scale, left plot) and $D$ (blue scale, right plot) of the
    least squares orbits obtained from these solutions. The red dots
    correspond to preliminary orbits that do not converge in the
    differential corrections scheme.}
  \label{fig:difCorTrueReal}
\end{figure}

Differential corrections were then applied to the best solutions (in
terms of $D$) of the two KI methods obtained with true linkages in the
real data.  The values of the $rms$ of the least squares orbits
($R_{LS}$) are shown in Figure \ref{fig:difCorTrueReal}.  More than
80\% of the \texttt{link2} preliminary solutions converge to a least
squares orbit but, due to the correlations in the observational
errors, the values of $R_{LS}$ are larger using real data than the
synthetic data set with 0.2$\arcsec$ error.

\begin{figure}[ht!]
  \centering
  \includegraphics[width=0.19\textwidth]{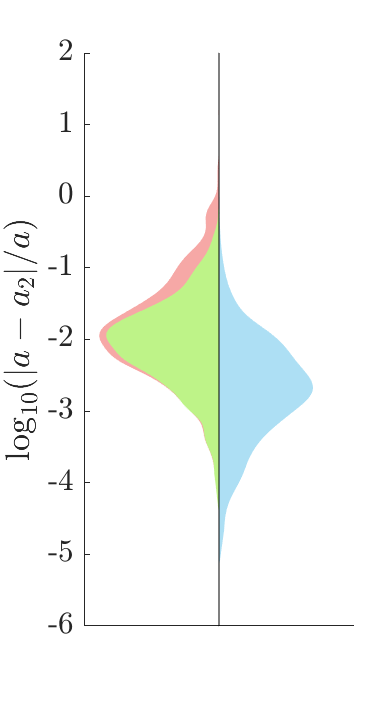}
  \includegraphics[width=0.19\textwidth]{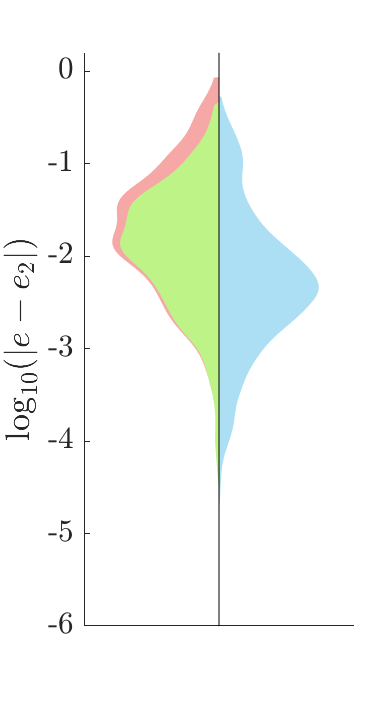}
  \includegraphics[width=0.19\textwidth]{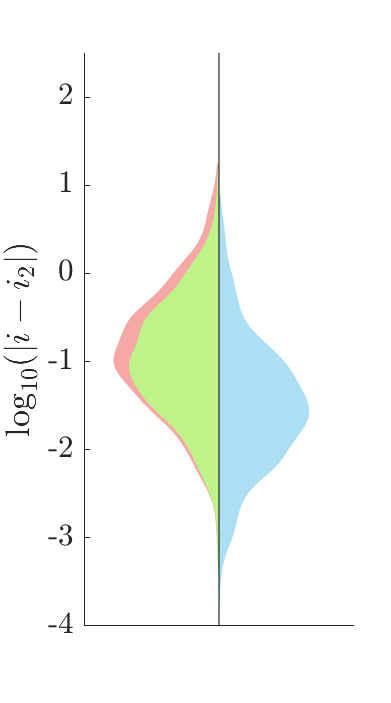}
  \includegraphics[width=0.19\textwidth]{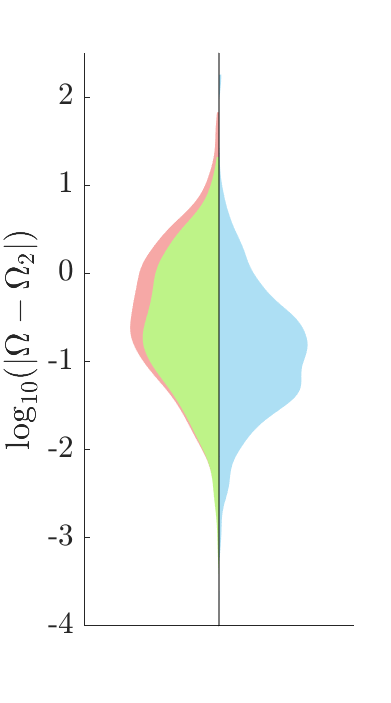}
  \includegraphics[width=0.19\textwidth]{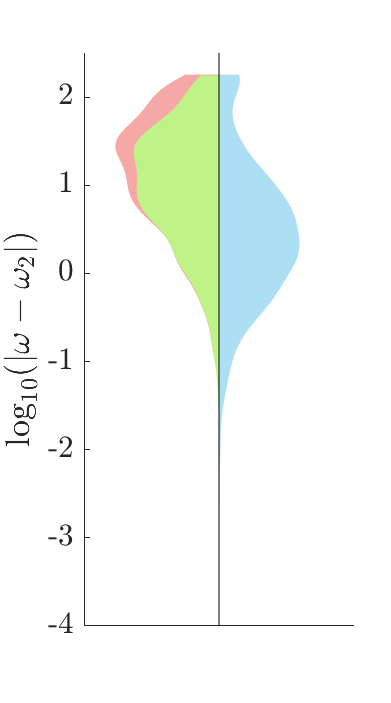}
  \caption{PDF of the logarithm of the error in $a$, $e$, $i$,
    $\Omega$, and $\omega$ (from left to right) for the solutions of
    \texttt{link2} that converge (green) and do not converge (red) to
    a least squares orbit using real data. In blue the same PDFs are
    shown for the least squares orbits.}
  \label{fig:lk2vsconvpdfReal}
\end{figure}

\begin{figure}[ht!]
  \centering
  \includegraphics[width=0.33\textwidth]{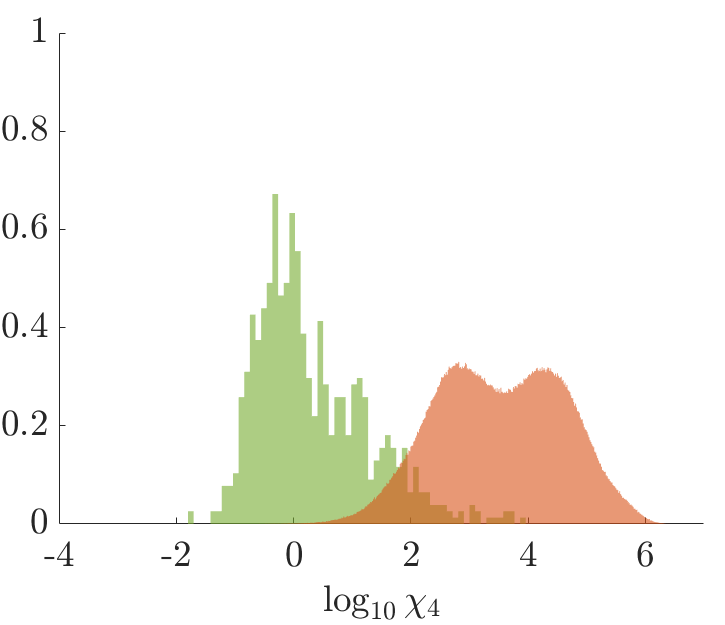}
  \includegraphics[width=0.33\textwidth]{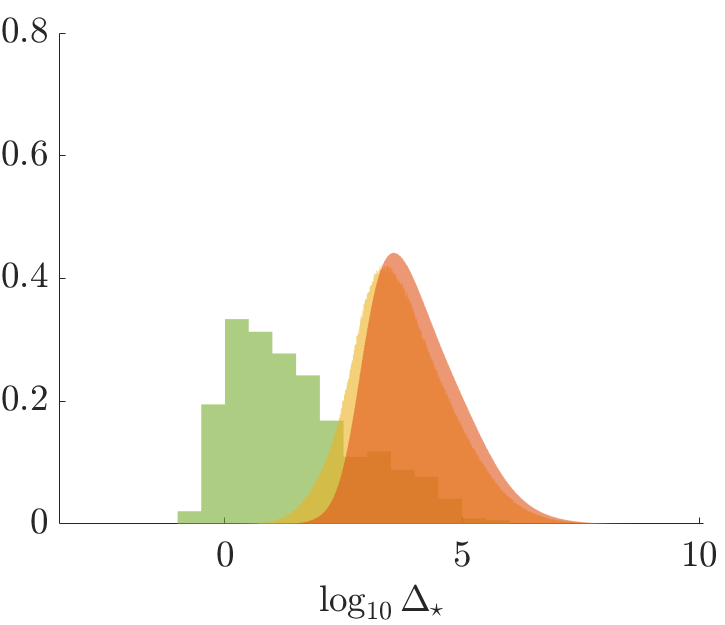}
  \caption{Left: PDF of the logarithm of $\chi_4$ for the true (green)
    and false (red) linkages of \texttt{link2} using real data.
    Right: PDF of the logarithm of $\Delta_\star$ for the true (green)
    linkages of \texttt{link3} and the false ones with two tracklets
    belonging to the same object (yellow) and each tracklet belonging
    to different objects (red) using real data.}
  \label{fig:TvFreal}
\end{figure}

As expected, the least squares orbits are better than the preliminary
ones, not only in terms of $D$ but also in the error of each orbital
element, see Figure \ref{fig:lk2vsconvpdfReal}. In this figure we
also observe that the orbits that do not converge are affected by
larger errors in the orbital elements.

\begin{figure}[ht!]
  \centering
  \includegraphics[width=0.66\textwidth]{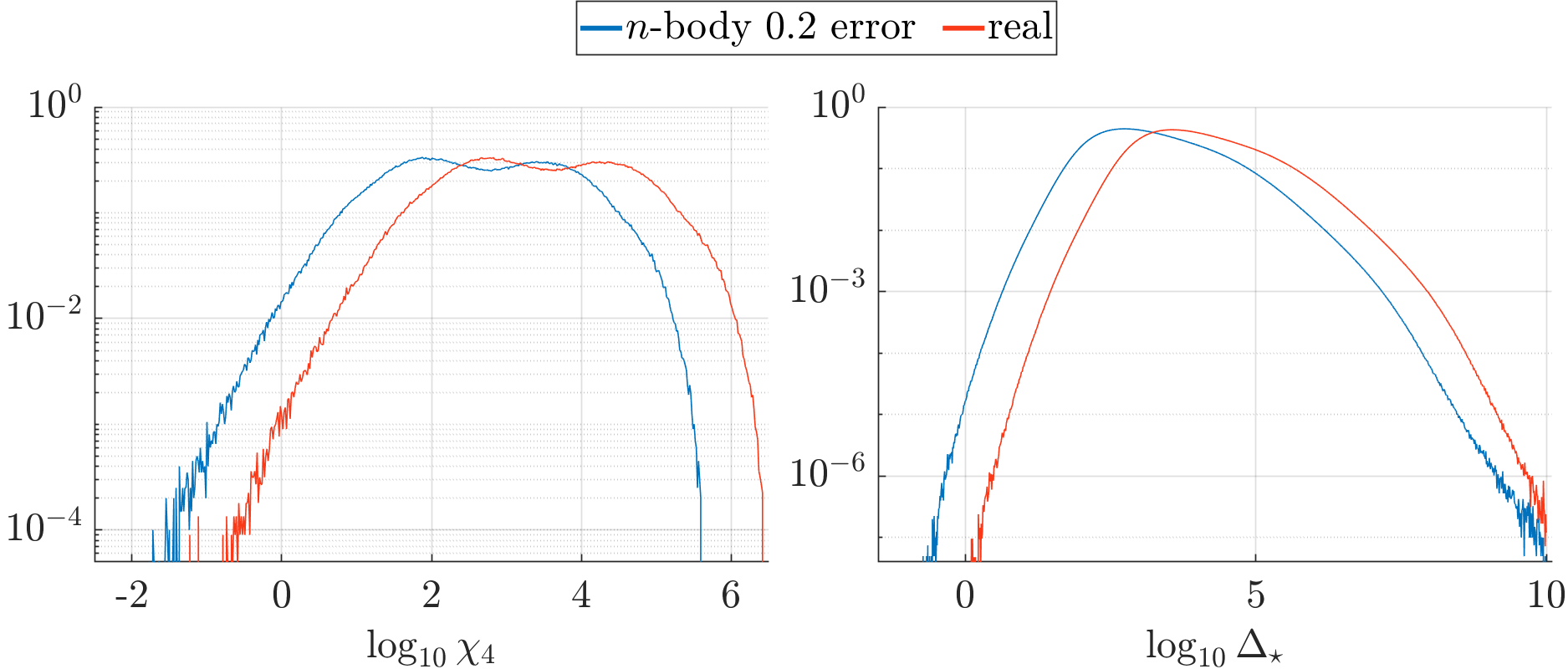}
  \caption{Left: PDF of the logarithm of $\chi_4$ for the false
    linkages of $\texttt{link2}$ using the synthetic data set
    generated by an $n$-body propagation with 0.2$\arcsec$ error
    (blue) and the real data (red). Right: the same for the PDF of the
    logarithm of $\Delta_\star$ with $\texttt{link3}$.}
  \label{fig:Fnbody02vsreal}
\end{figure}

\begin{figure}[ht!]
  \centering
  \includegraphics[width=0.33\textwidth]{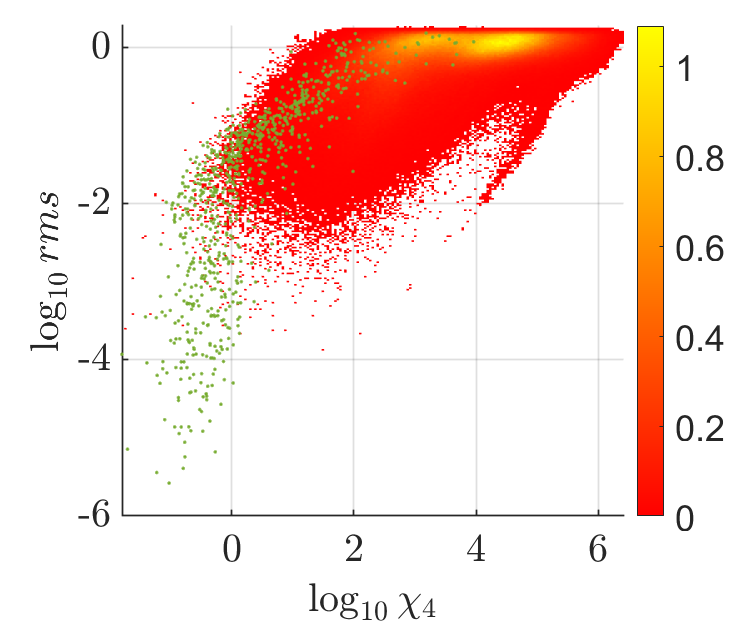}
  \includegraphics[width=0.33\textwidth]{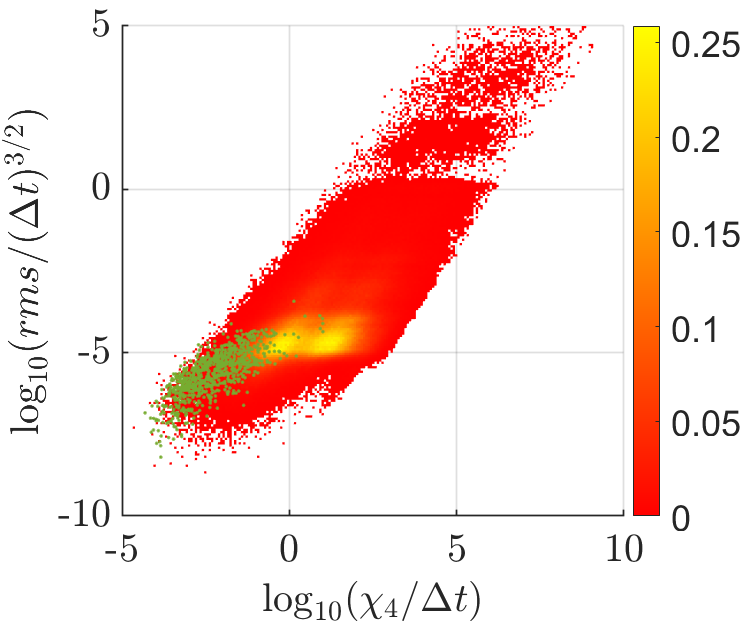}
  \caption{
    Same as in Figure \ref{fig:rms-chi4-syn} but for real
    observations.}
  \label{fig:rms-chi4-real}
\end{figure}

Next, we investigate the ability of the KI methods to recover true
linkages when applied to real observations. For this purpose, we apply
the same test that we performed with the synthetic data sets (see
Section~\ref{sec:TLinks}). In Figure \ref{fig:TvFreal} we show the PDF
of the $\chi_4$ and $\Delta_\star$ norms for true and false linkages.
These distributions are almost identical to those obtained for
synthetic observations with 0.2$\arcsec$ error (see for comparison the
corresponding plots in Figures \ref{fig:pdfLK2truevsfalse} and
\ref{fig:pdfLK3truevsfalse}), but they all have a small positive
displacement in the horizontal axis.

\begin{figure}[ht!]
  \centering
  \includegraphics[width=0.33\textwidth]{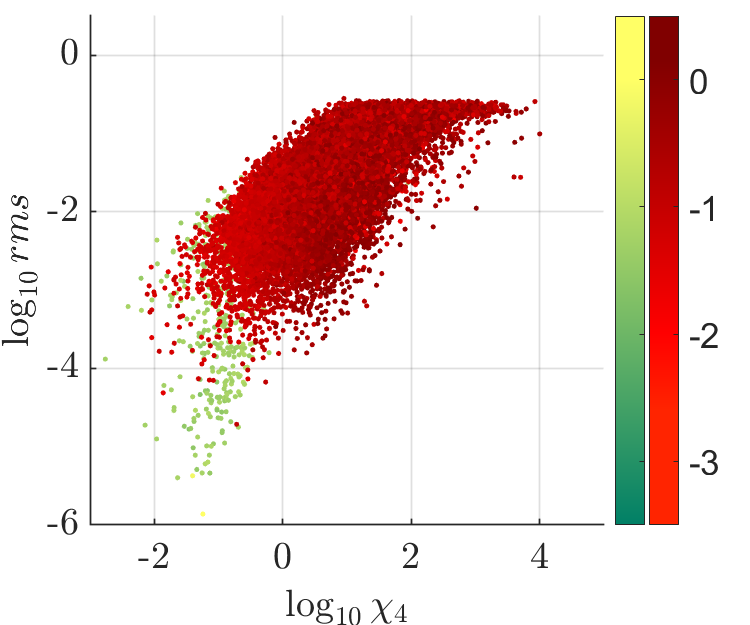}
  \includegraphics[width=0.33\textwidth]{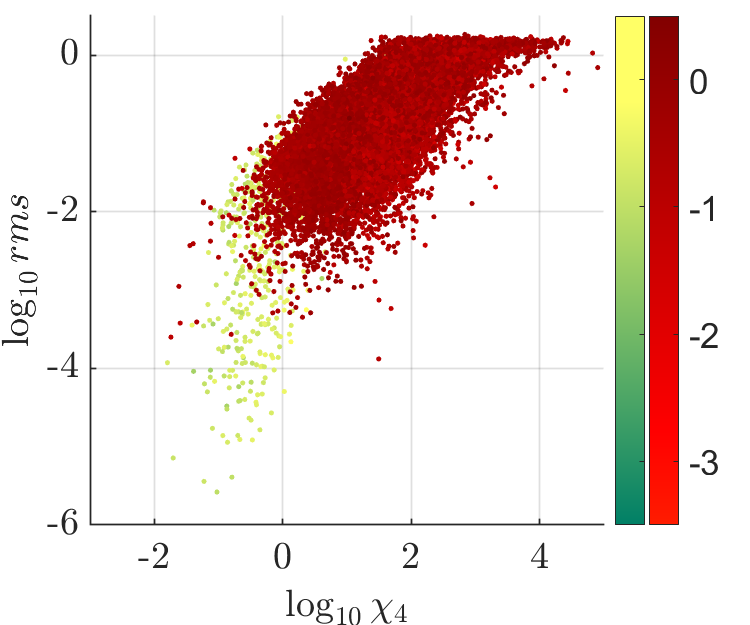}
  \caption{$rms$ vs $\chi_4$ in a log-log plot for the solutions of
    \texttt{link2} and logarithmic values of $R_{LS}$ of the least
    squares orbits obtained from the true (green scale) and false (red
    scale) solutions using synthetic observations generated by an
    $n$-body propagation with 0.2$\arcsec$ error (left) and real
    observations (right).}
  \label{fig:difCorReal}
\end{figure}

This phenomenon can also be observed from the comparison of Figures
\ref{fig:rms-chi4-real} and \ref{fig:rms-chi4-syn} and from Figure
\ref{fig:difCorReal}.

We emphasise that in this case near the 99\% of false linkages do not
converge to a LS orbit when we apply the differential corrections. We
also note that all the preliminary solutions with $\log_{10}\chi_4$
greater than 5 do not converge.

In conclusion, when dealing with real observations we can set the
thresholds of $\chi_4$ and $\Delta_\star$ by properly increasing the
values of these two norms previously found with the synthetic
populations (see Figure \ref{fig:Fnbody02vsreal}).

\section{Application to the ITF}
\label{sec:ITF}

One of our future goals is to apply the KI methods to the MPC's ITF to
identify tracklets that can be associated with known objects and to
discover unknown objects with tracklets that have not been linked
together. In this section we estimate bounds on the technique's
efficiency with some assumptions on the data's characteristics and the
method's application.

First, we assume that the application to the ITF will use
\texttt{link2} since it has a higher efficiency at recovering true
linkages than \texttt{link3} (98\% vs. 84\% for real observations) and
the quality of its solutions is good.

\begin{figure}[ht!]
  \centering
  \includegraphics[width=0.66\textwidth]{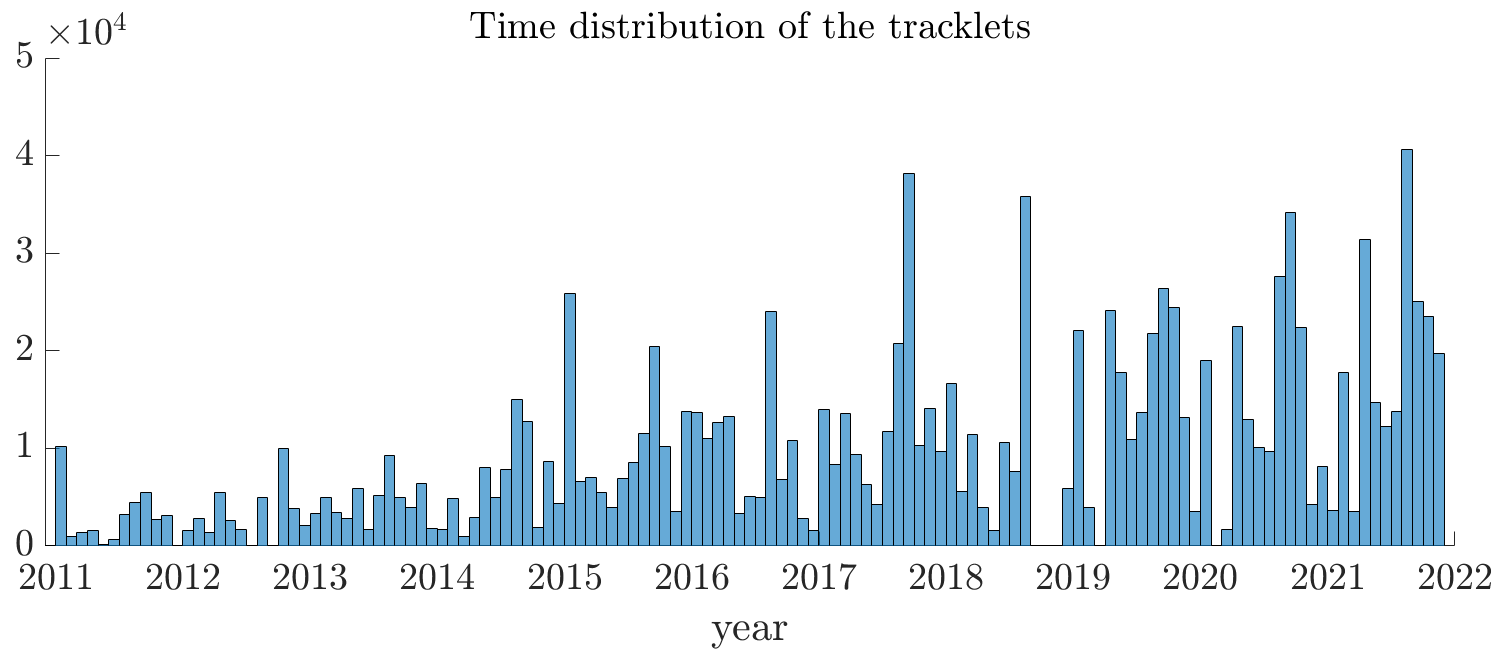}
  \caption{Distribution of the average time of observation of
    Pan-STARRS1 tracklets containing $\ge3$ detections in the ITF.}
  \label{fig:trackletF51}
\end{figure}

Second, for the purpose of this calculation, and to ensure that the
data has consistent properties, e.g. astrometric and photometric
accuracy, we only consider ITF tracklets containing $\ge3$ detections
contributed by Pan-STARRS1 which dominates the ITF at almost 50\% of
the total number of tracklets.  This data set contains $N=1,252,187$
tracklets so it is about 760 times larger than the synthetic and real
data sets used with \texttt{link2} in the previous sections and its
time distribution can be observed in Figure \ref{fig:trackletF51}.
The increasing number of unlinked tracklets as a function of time is
likely due to Pan-STARRS1 modifying its survey strategy and
incremental improvements in system operations
\citep{chambers2019panstarrs1}.

Since the computational cost of a full numerical exploration is
$\mathcal{O}(N^2)$, where $N$ is the number of tracklets in the data
set, the time of computation and the number of (true and false)
solutions of \texttt{link2} will increase by a factor $\approx$
580,000 relative to the cost of exploring our small subset of real
Pan-STARRS1 ITF data (see Table \ref{tab:solutionsITF}).
\begin{table}[ht!]
  \centering
  \begin{tabular}{lrr}
    & \multicolumn{1}{l}{real data} & ITF F51 data \phantom{}  \\ \hline
    Number of solutions & 1,861,785 & $\approx$1.08$\,\cdot10^{12}$  \\
  \end{tabular}
  \caption{Total number of solutions (true and false, including
    multiple solutions, without setting a threshold for $\chi_4$)
    generated by {\texttt{link2}} with the real data set of 822
    objects used in this work (Section~\ref{sec:realdata}) and the
    expected total number of solutions with Pan-STARRS1 (F51)
    tracklets extracted from the ITF.}
  \label{tab:solutionsITF}
\end{table}

\begin{figure}[ht!]
  \centering
  \includegraphics[width=0.66\textwidth]{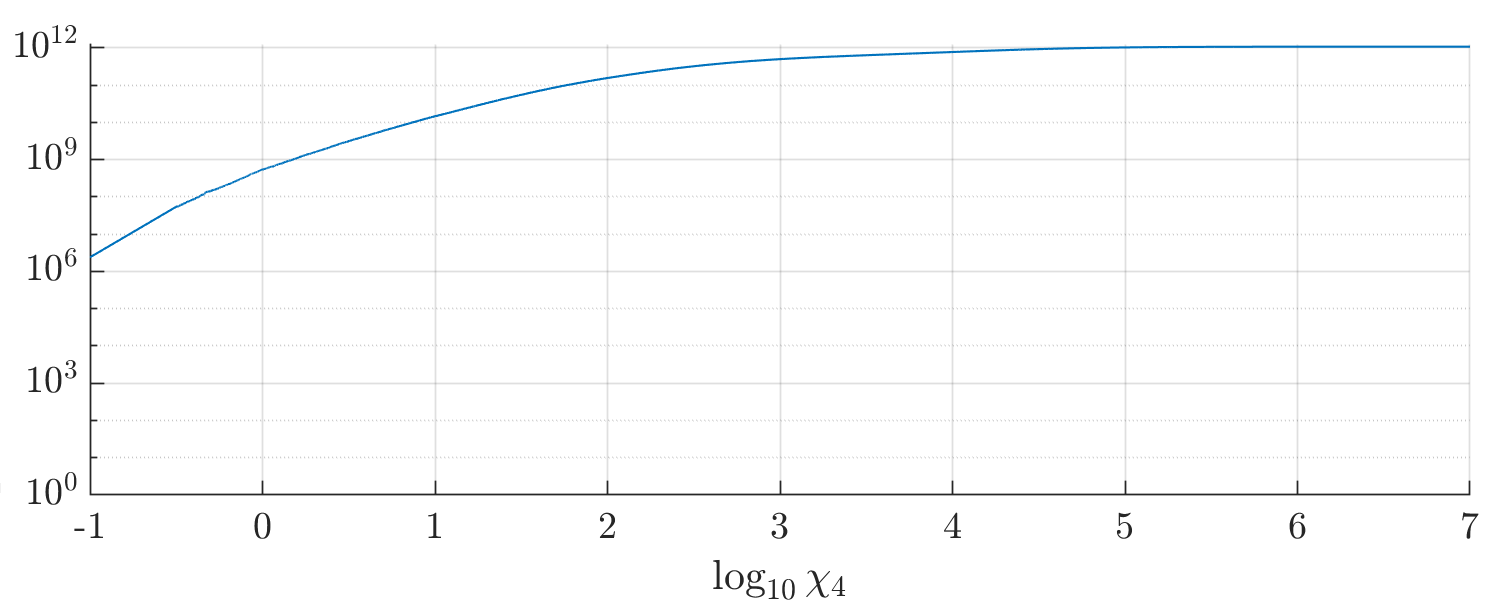}
  \caption{Expected number of solutions as a function of the threshold
    of the logarithm of $\chi_4$ for the set of Pan-STARRS1 (F51)
    tracklets contained in the ITF.}
  \label{fig:expNumbtrackletF51}
\end{figure}

The identification of a new object in the ITF will require linking
$\ge4$ tracklets. The combinatorics of linking tracklets within the
ITF can be expressed as a graph $G=G(V,E)$ where $V$, the set of
vertices, corresponds to the set of tracklets (i.e. $V =
\{1,2,...,N\}$) and $E$, the set of edges, corresponds to the
linkages, i.e. we will consider $e_{ij}\in E$ with $i,j\in V$ if and
only if we obtain a linkage between the tracklets $i$ and $j$ using
\texttt{link2}.  Thus, we need to estimate the number of 4-connected
sub-graphs of $G$.

To that end we define the random variable $X_{ij}$ as a function of
the threshold value $\chi_4^*$ of $\chi_4$:
\[
X_{ij}(\chi_4^*) = \left\{
\begin{aligned}
  & 1 \text{if we obtain at least one solution for the tracklets}  \\[-1ex] 
  & \phantom{1} \text{$i$ and $j$ with $\chi_4\leq\chi_4^*$,}  \\
  & 0 \text{otherwise,}
\end{aligned}
\right.
\]
so $X_{ij} = 1$ if $e_{ij}\in E$, and $X_{ij} = 0$ otherwise.

\begin{figure}[ht!]
  \centering
  \includegraphics[width=0.66\textwidth]{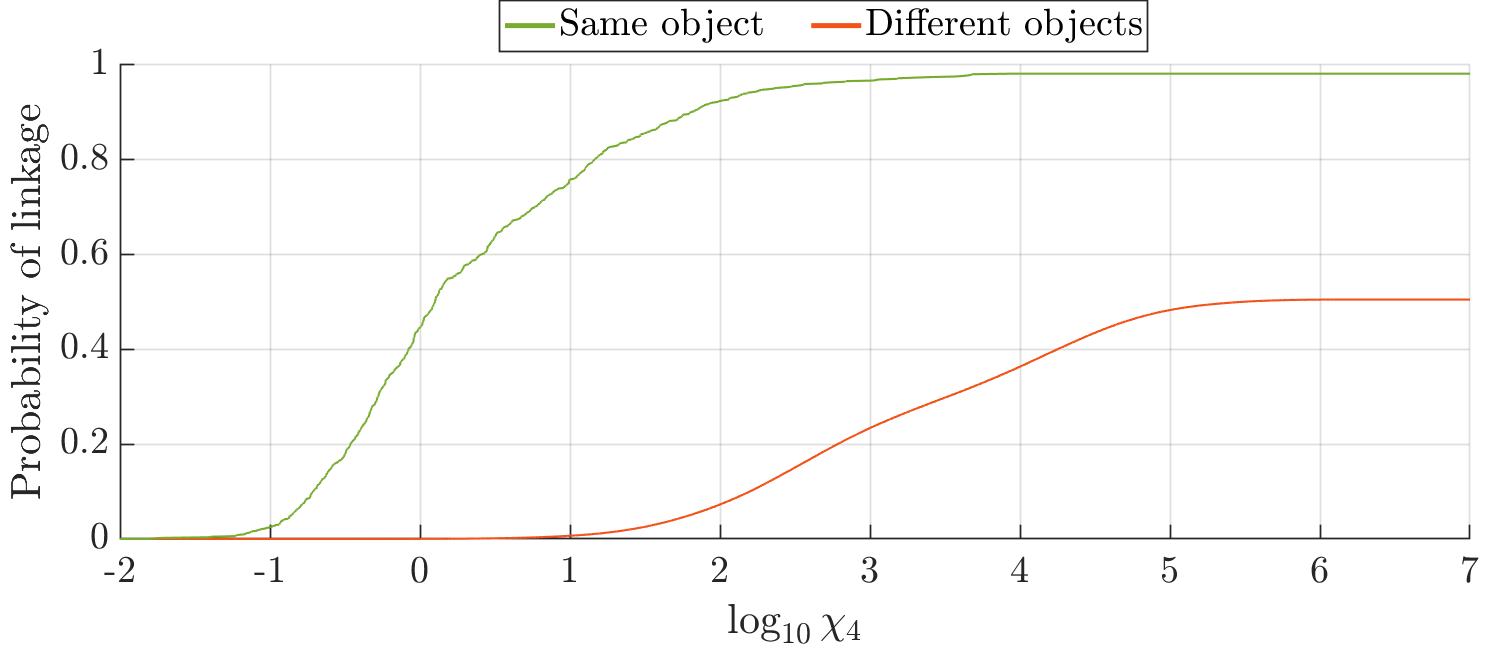}
  \caption{Probability of obtaining a preliminary orbit with
    \texttt{link2} if two tracklets belong to the same object (green)
    or not (red), as a function of the threshold value of $\chi_4$.}
  \label{fig:probabilityLinkages}
\end{figure}

From the previous analysis, we can estimate the probability
$P(X_{ij}=1)$ as a function of $\chi_4^*$, taking also into account
whether the tracklets $i$ and $j$ belong or not to the same
object. These results can be observed in
Figure~\ref{fig:probabilityLinkages}. We note that the probability to
obtain a linkage with 2 tracklets belonging to different objects gets
very close to $0.5$ as $\chi_4$ increases.

It is important to note that we do not know the distribution of the
number of tracklets per object in the ITF and therefore we cannot
determine the number of 4-connected sub-graphs that will be found.  To
set a lower bound on the problem we assume that there is a single
object with 4 tracklets in the ITF and the rest of the $N-4$ tracklets
belong to $N-4$ different objects.  Furthermore, we assume that the
linkages are independent i.e. there is no correlation between the
\texttt{link2} solutions of pairs or tracklets belonging to the same
object.

\begin{figure}[ht!]
  \centering
  \includegraphics[width=0.66\textwidth]{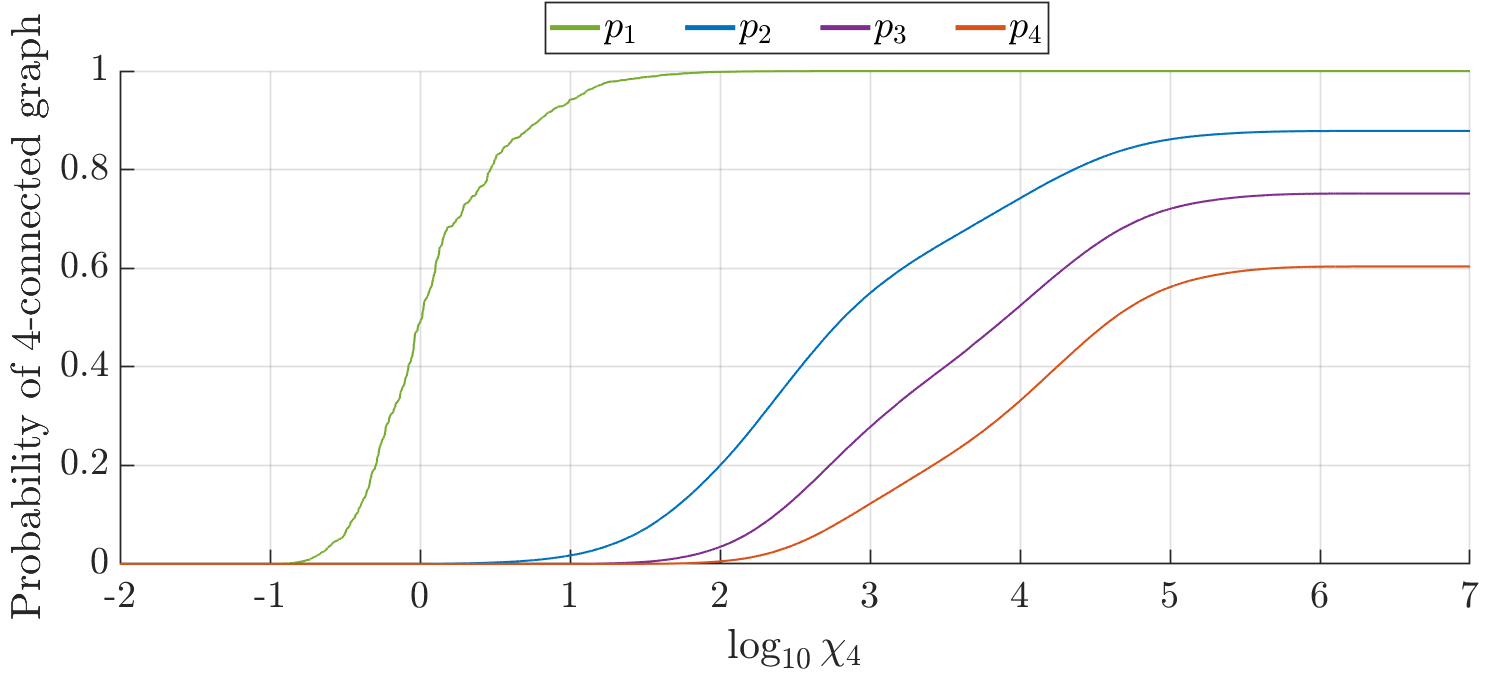}
  \caption{Probability $p_k$ of identifying a 4-connected sub-graph
    given 4 tracklets that belong to $k=1,...,4$ different objects as
    a function of the $\chi_4$ threshold.}
  \label{fig:4-Connected-probability}
\end{figure}

To calculate the number of 4-connected sub-graphs that must be tested,
let us define $p_k$ as the probability that 4 tracklets belonging to
$k=1,...,4$ different objects form a 4-connected subgraph.  The values
of $p_k$ as a function of $\chi_4$ are shown in Figure
\ref{fig:4-Connected-probability} (see more details in Appendix
\ref{app:graph}).  As the $\chi_4$ threshold is loosened the
probability of identifying the correctly linked 4-connected sub-graph
increases quickly around $\log_{10}\chi_4=0$ and by
$\log_{10}\chi_4=1$ there is nearly 100\% probability that the set of
4 tracklets will be identified.  On the other hand, the probability
that 1 of the 4 tracklets will be an unrelated interloper begins to
increase around $\log_{10}\chi_4=1$ and by $\log_{10}\chi_4\approx3$
the probability that all 4 tracklets are unrelated is $>10$\%.

In this way, the expected number ($N_4$) of 4-connected sub-graphs in
the data set as a function of $\chi_4$ is given by (see Appendix
\ref{app:graph})
\begin{equation}\label{eq:N4}
  N_4(\chi_4) = p_1 + 4(N-4)p_2 + 6\binom{N-4}{2}p_3 
  + \left[4\binom{N-4}{3}+\binom{N-4}{4}\right]p_4.
\end{equation}

\begin{figure}[ht!]
  \centering
  \includegraphics[width=0.66\textwidth]{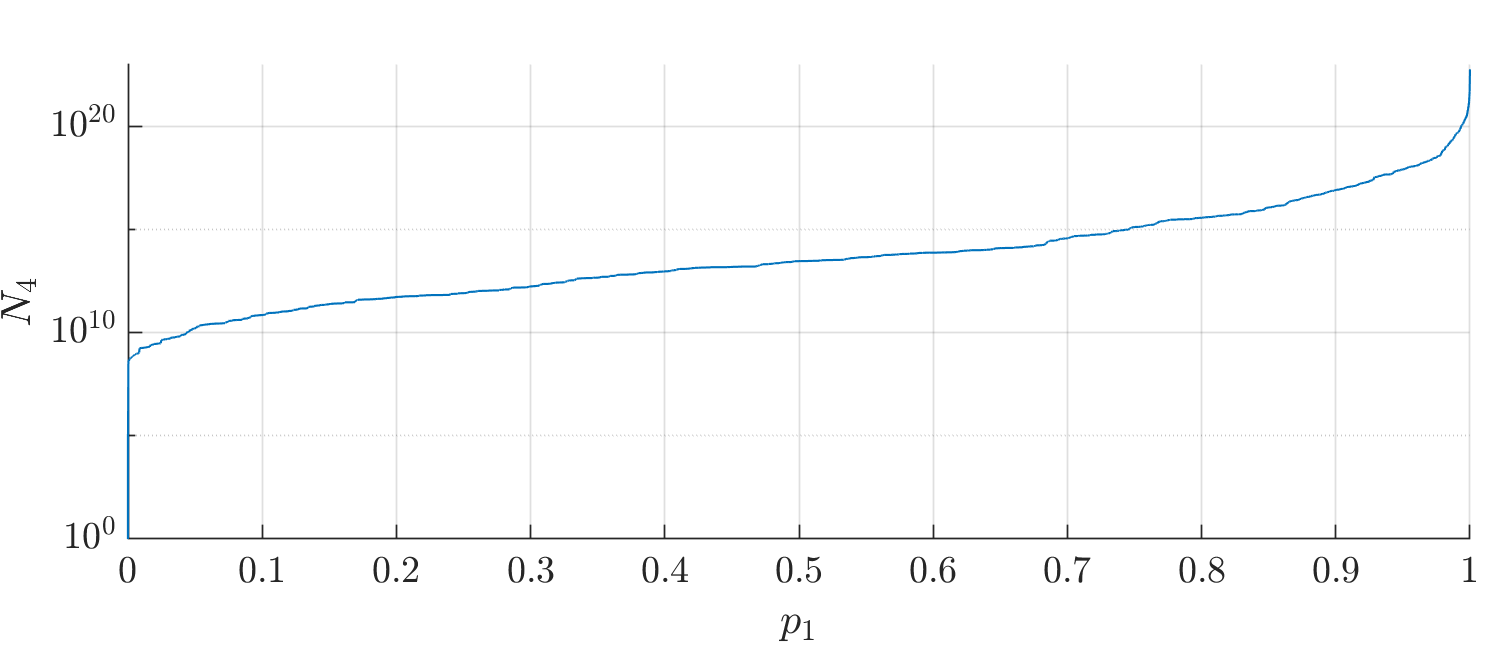}
  \caption{Expected number of 4-connected sub-graphs that must be
    tested as a function of the probability, $p_1$, of identifying the
    correctly linked set of 4 tracklets.}
  \label{fig:N4-vs-probability2}
\end{figure}

Since each $p_k$ is an increasing function of $\chi_4$ we can
calculate the probability of detecting the single set of 4 tracklets
corresponding to the same object as a function of the number of sets
of 4 tracklets that must be tested (i.e. the number of 4-connected
sub-graphs $N_4$), see Figure~\ref{fig:N4-vs-probability2}.

Since we expect that the ITF contains more than one object that has at
least 4 tracklets, Figure \ref{fig:N4-vs-probability2} provides a
lower bound on the number of 4-tracklets sets that must be tested to
recover the desired fraction of unidentified objects (i.e. the value
of $p_1$, in our simple case).

These results may seem discouraging because they suggest that we will
need to test more than $10^{15}$ sets of 4-tracklets to recover at
least 80\% of the objects and that number of differentially corrected
orbit computations would be challenging.  Thus, we will need to invoke
extra conditions on the 4-tracklets before attempting a least squares
orbit to drastically reduce the computational problem.  We expect that
the \texttt{link2} preliminary orbit solutions will provide the
necessary reduction by requiring that pairs of tracklets in the set of
connected tracklets have similar Keplerian integrals.

We will further develop this technique in its application to the ITF
in our next paper, but it is clear that achieving detection
efficiencies of $>90$\% for objects with just 4-tracklets in the ITF
is a challenging task.

\section{Conclusions}

Two linkage methods for initial orbit determination, named
\texttt{link2} and \texttt{link3}, have been analysed using synthetic
and real data, with the goal of understanding whether they can be
efficiently applied to large repositories of unlinked detections, such
as the MPC's ITF. The low computational cost of these algorithms make
them promising for this purpose.

The results obtained with synthetic data generated with astrometric
errors typical of modern wide-field asteroids surveys are good: the
percentage of recovered true linkages is high, and the preliminary
orbits are close to the real ones.  In these two aspects
\texttt{link2} is better than \texttt{link3}. Furthermore, the
orbital plane is quite well determined by both methods.  On the other
hand, the percentage and the quality of the solutions decrease for
larger astrometric error.

Some indicators to estimate the quality of the preliminary solutions
have been studied. These indicators, $\chi_4$ for \texttt{link2},
$\Delta_*$ for \texttt{link3} and the \emph{rms} of the orbit for both
methods, show a significant correlation with the quality of the
preliminary orbits, which was quantified with the $D$-criterion.  By
setting suitable thresholds for these indicators a large fraction of
the false linkages, which are unavoidably produced with both methods,
can be discarded, without losing too many true ones. This operation is
less effective as the error increases.

Using some simple assumptions we have seen that a preliminary
exploration of the ITF is computationally feasible with a procedure
relying on \texttt{link2}.  For this reason, we believe that
\texttt{link2} is a good method to carry out a complete exploration of
the ITF observations made by Pan-STARRS1.

\section*{Acknowledgments}
This work was partially supported through the H2020 MSCA ETN
Stardust-Reloaded, Grant Agreement Number 813644.  OR acknowledges the
Spanish MINECO/FEDER grant PGC2018-100928-B-I00.  GFG and GB also
acknowledge the project MIUR-PRIN 20178CJA2B ``New frontiers of
Celestial Mechanics: theory and applications" and the GNFM-INdAM
(Gruppo Nazionale per la Fisica Matematica).  Part of this work was
performed during RJ's visits in Pisa.

\section*{Data Availability}

The data underlying this article will be shared on reasonable request
to the corresponding author.

\printbibliography


\setcounter{table}{0} \renewcommand{\thetable}{A.\arabic{table}}
\setcounter{section}{0} \renewcommand{\thesection}{A.}
\setcounter{subsection}{0} \renewcommand{\thesubsection}{A.\arabic{subsection}}

\appendix

\section{Computation of the probabilities associated to the 4-connected subgraphs}
\label{app:graph}

We made the assumption that the \texttt{link2} solutions are
independent in our estimation of the expected number of 4-connected
sub-graphs that will be obtained from an observation data set
containing $N-3$ objects where 1 object has 4 tracklets and $N-4$
objects have only 1 tracklet. This means that we assumed that the
probability of linking two tracklets depends only on whether they
belong to the same object. Thus, we let $p$ be the probability of
linking two tracklets that belong to the same object and we denote by
$q$ the probability of linking two tracklets that belong to different
objects. The values of $p$ and $q$ as a function of the threshold
value of $\chi_4$ are displayed in Figure
\ref{fig:probabilityLinkages}.

\begin{figure}[ht!]
  \centering
  \includegraphics[width=0.66\textwidth]{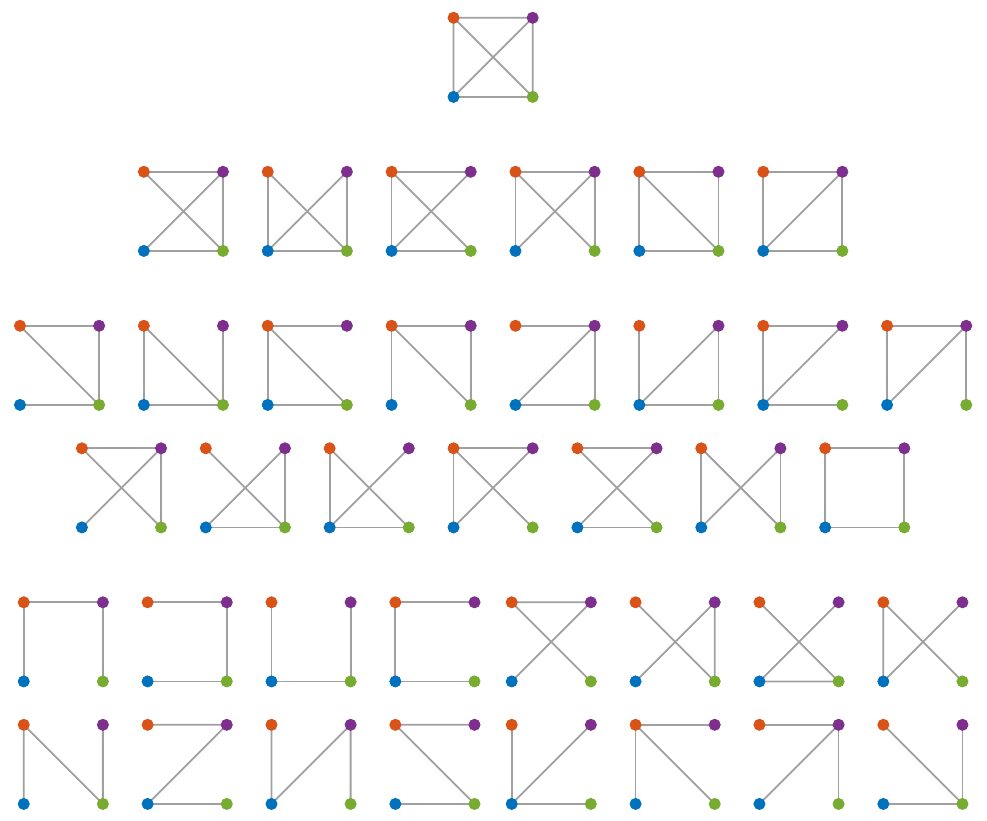}
  \vspace{3mm}
  \caption{The 38 possible 4-connected graphs.}
  \label{fig:4congraf}
\end{figure}

Given 4 vertices, there are 38 possible 4-connected graphs as shown in
Figure \ref{fig:4congraf}. From this figure we can compute the
probability $p_k$ to have a 4-connected sub-graph if the tracklets
belong to $k$ objects for $k=1,2,3,4$.  Note that, due to our simple
assumptions on the data set, the only way to have 4 tracklets
belonging to 2 objects is that 3 tracklets belong to one of them.  In
this way, these probabilities are:
\[
\begin{aligned}
  p_1 &= p^6 + 6p^5(1-p) + 15p^4(1-p)^2 + 16p^3(1-p)^3,\\[1.5ex]
  p_2 &=  p^3q^3 
  + 3p^3q^2(1-q) + 3p^2(1-p)q^3  
  + 3p^3q(1-q)^2 \\
  & \quad 
  + 9p^2(1-p)q^2(1-q)
  + 3p(1-p)^2q^3
  + (1-p)^3q^3 \\
  & \quad 
  + 6p(1-p)^2q^2(1-q)
  + 9p^2(1-p)q(1-q)^2,\\[1.5ex] 
  p_3  & =  pq^5 
  + 5pq^4(1-q) + (1-p)q^5  
  + 5(1-p)q^4(1-q) \\
  & \quad
  + 10pq^3(1-q)^2
  + 8pq^2(1-q)^3
  + 8(1-p)q^3(1-q)^2,\\[1.5ex]
  p_4 & = q^6 + 6q^5(1-q) + 15q^4(1-q)^2 + 16q^3(1-q)^3.\\
\end{aligned}
\]

Finally, formula \eqref{eq:N4} comes from the property of the
binomials
\[
\binom{N}{M} = \sum_{k=0}^M \binom{M}{M-k}\binom{N-M}{k},
\]
for the particular case of $M=4$ using the probabilities computed
previously:
\[
\begin{aligned}
  N_4(\chi_4) &= 
  \binom{4}{4}\binom{N-4}{0}p_1 + 
  \binom{4}{3}\binom{N-4}{1}p_2 +  
  \binom{4}{2}\binom{N-4}{2}p_3\\ 
  & \quad +
  \binom{4}{1}\binom{N-4}{3}p_4 + 
  \binom{4}{0}\binom{N-4}{4}p_4\\[0.5ex]
  & = p_1 + 4(N-4)p_2 + 6\binom{N-4}{2}p_3 
  + \left[4\binom{N-4}{3}+\binom{N-4}{4}\right]p_4.
\end{aligned}
\]



\end{document}